\documentclass[12pt, draftclsnofoot, onecolumn]{IEEEtran}
\usepackage{amsfonts,enumerate,epsfig,cite,array,multirow,graphicx,amsmath,amsthm,ltablex,tabularx,setspace,arydshln,amssymb,multirow,subfigure}
\usepackage{tikz}
\usepackage{schemabloc,tikz}
\usetikzlibrary{arrows}
\usepackage[colorlinks]{hyperref}
\usepackage{acronym} % Use \Ac{qam} or \ac{qam}
\acrodef{iot}[IoT]{internet-of-things}
\acrodef{6g}[6G]{sixth-generation}
\acrodef{backcom}[BackCom]{Backscatter communication}
\acrodef{pls}[PLS]{Physical layer security}
\acrodef{eh}[EH]{energy-harvesting}
\acrodef{sots}[SOTS]{sub-optimal tag selection}
\acrodef{mets}[METS]{minimal eavesdropping tag selection} \acrodef{ots}[OTS]{optimal tag selection}
\acrodef{rts}[RTS]{random tag selection}
\acrodef{pdf}[PDF]{probability density function}
\acrodef{pdfs}[PDFs]{probability density functions}
\acrodef{cdf}[CDF]{cumulative distribution function}
\acrodef{mgf}[MGF]{moment generating function}
\acrodef{op}[OP]{outage probability}
\acrodef{sop}[SOP]{secrecy outage probability}
\acrodef{ip}[IP]{intercept probability} 
\acrodef{cc}[CC]{channel capacity}
\acrodef{csi}[CSI]{channel state information}
\acrodef{se}[SE]{spectral efficiency}
\acrodef{aser}[ASER]{average symbol error rate}
\acrodef{aber}[ABER]{average bit error rate}
\acrodef{e2e}[e2e]{end-to-end}
\acrodef{snr}[SNR]{signal-to-noise ratio}
\acrodef{awgn}[AWGN]{additive white Gaussian noise}
\acrodef{acc}[ACC]{average channel capacity}
\acrodef{do}[DO]{diversity order}
\acrodef{cc}[CC]{channel capacity}
\acrodef{esc}[ESC]{ergodic secrecy capacity}
\acrodef{sr}[SR]{secrecy rate}
\acrodef{inid}[i.n.i.d.]{independent but non-identically distributed }
\acrodef{iid}[i.i.d.]{independent and identically distributed }
\acrodef{awgn}[AWGN]{additive white Gaussian noise}

\begin{document}
	\title{Secrecy Analysis of Energy-Harvesting Backscatter Communications with Tag Selection in Nakagami-$m$ Fading}
	\author{Mohammad Nafees, \textit{Student Member, IEEE,} Dharmendra Dixit, \textit{Member, IEEE},   and Arvind Kumar,~\IEEEmembership{Member,~IEEE}
		\thanks{The authors are with the Electronics and Communication Engineering Department, Motilal Nehru National Institute of Technology Allahabad, Prayagraj, 211004, India
			e-mail: (nafeesiiit@gmail.com (arvindk, dharmendradixit))@mnnit.ac.in}
	}
\maketitle
\begin{abstract} Backscatter communication is an energy-efficient technique that enables sustainable wireless connectivity with a minimal environmental impact. In this paper, the secrecy performance of practical non-linear energy-harvesting backscatter communications with various tag selection schemes is analyzed in Nakagami-$m$ fading channels. We consider four tag selection schemes: sub-optimal, minimal eavesdropping, optimal, and random tag selection. Closed-form expressions for \ac{sop} and \ac{ip} are derived for each scheme, along with asymptotic expressions to provide deeper insights. The impact of system and fading parameters on \ac{sop} and \ac{ip} is investigated, and simulation results are presented to validate the accuracy of the analytical expressions.
\end{abstract}
 % Reset All the Acronym counts
\acresetall
% 	%%%%%%%%%%%% 	
\section{Introduction}
\IEEEPARstart{A}{s} multi-device \ac{iot} ecosystems continue to expand, future wireless technologies such as sixth-generation (6G) networks face significant challenges in spectrum optimization and power consumption, particularly in low-cost, low-power, and battery-free systems \cite{kludze2024frequency,van2018ambient,jia2023secrecy}. \ac{backcom} is an emerging technology that addresses these challenges by enabling devices to transmit information through the modulation of ambient radio signals without requiring dedicated power sources. This innovative method has garnered significant attention in \ac{iot} applications due to its minimal power requirements and potential for widespread deployment. However, \ac{backcom}’s relatively simple transmission structure makes it vulnerable to passive eavesdropping, where adversaries may intercept and decode transmitted data. Furthermore, the shared nature of the unlicensed wireless spectrum introduces concerns regarding interference and unauthorized access. Therefore, securing BackCom systems is essential. \ac{pls} is a promising solution to ensure secure communication between authorized nodes with minimal complexity and high efficiency \cite{jia2023secrecy,illi2023physical,hamamreh2018classifications}. By leveraging physical characteristics like noise and channel fluctuations, \ac{pls} enhances security, making it a promising approach for safeguarding \ac{backcom} systems.
 
%   BackCom, which uses a dedicated signal resource such as a radio frequency identification (RFID) system, was thoroughly researched in the first few decades after the invention of BackCom\cite{11},\cite{12}. When equipped with a dedicated signal source, the backscatter transmitter can be placed closer to the signal source, thus reducing the associated path loss. Many methods have been developed to reduce backscatter path loss between transmitter and receiver and to enhance system performance, including channel coding\cite{13},\cite{14}, multiple-input multiple-output (MIMO) techniques\cite{15}, collision avoidance protocols\cite{16}, wireless channel simulation\cite{17} etc. Generally, BackCom with a dedicated signal source is used in many practical applications, such as commerce, strategic positioning, etc. This setup can also reduce the cost and size of the system. However, despite these advantages, BackCom still hard to meet the demands of data-intensive communication with a dedicated signal source\cite{18}.
  
%      %%%%%%%%%%%%%%%   	
\subsection{Related Works}
Several studies in the literature have investigated the secrecy performance of \ac{backcom} systems. 
%%%% 3-10
%%% [3] SOP Rayeligh
In \cite{liu2021secrecy}, the \ac{sop} for a passive \ac{backcom} system with multiple tags is derived for Rayleigh fading channels. 
%%% [4] SR without fading (verify)
The \ac{sr} of a single-reader, single-tag model is analyzed across various scenarios using analytical methods in \cite{saad2014physical}.
%%% [5] SOP, Correlated Rayliegh
In \cite{zhang2019secure}, the authors are obtained the exact and asymptotic \ac{sop} expressions for a multi-tag \ac{backcom} system in correlated Rayleigh fading channels.
%%% [6]
In \cite{liu2022secrecy}, analytical expressions for \ac{esc} and \ac{sop} are derived for a multi-tag self-powered \ac{backcom} network using a proposed tag selection strategy in Nakagami-$m$ fading. 
%%% [7] X
% The paper investigates two strategies—jamming and cooperation—for enhancing physical layer security in multi-tag ambient backscatter communications, proposing scheduling methods that either generate artificial noise to disrupt eavesdroppers or enable cooperative transmission to assist the receiver, and evaluates both strategies through simulations to provide guidelines for their optimal use based on system parameters.
%%% [8]
In \cite{muratkar2021physical} the exact and asymptotic \ac{sop}s are presented for multi-tag ambient \ac{backcom} systems, considering motion and imperfect channel estimation in Rayleigh fading.
%%% [9]
%%% [10]]
In \cite{wang2017wirelessly}, the bit error rate performance of ambient \ac{backcom} systems in Rayleigh fading was investigated through theoretical and simulation results.
%%%% 11-20
%%% [11]
%%% [12]
%%% 
%%% [13]
In \cite{khan2024secrecy}, expressions for the \ac{sop} and \ac{ip} of a multi-tag \ac{backcom} network with a practical non-linear \ac{eh} model are derived in Rayleigh fading for four tag selection schemes: \ac{sots}, \ac{mets}, \ac{ots}, and \ac{rts}. Based on this literature survey and to the best of the author's knowledge, the secrecy performance of \ac{backcom} systems incorporating practical non-linear \ac{eh} models and tag selection schemes (\ac{sots}, \ac{mets}, \ac{ots}, \ac{rts}) has not been explored in the context of versatile Nakagami-$m$ fading model.
%%% [14]
%%% [15]
 
%%%%%%%==Motivation and Contributions
\subsection{Motivation and Contributions}
 Nakagami-$m$ fading distribution is widely used to model various wireless propagation scenarios, such as urban and indoor environments, due to its flexibility in characterizing different fading conditions \cite{SimonBookDC}. This adaptability makes it crucial for designing robust \ac{backcom} systems, ensuring enhanced performance and reliability. Despite the practical significance of Nakagami-$m$ fading, the secrecy performance of \ac{backcom} systems with practical non-linear \ac{eh} models and tag selection schemes remains unexplored. In this work, we investigate the secrecy performance of a multi-tag \ac{backcom} network under Nakagami-$m$ fading for four tag selection protocols: \ac{sots}, \ac{mets}, \ac{ots}, and \ac{rts}.
 
 The key contributions of this work can be summarized as follows:

 \begin{enumerate}[i)]
     \item We derive new closed-form analytical expressions for \ac{sop} and \ac{ip} for each selection scheme in Nakagami-$m$ fading channels.
     \item To provide deeper insights, we also derive asymptotic expressions for \ac{sop} and \ac{ip} for each selection scheme.
     \item We examine the impact of variations in the number of tags, distances, and threshold rates on the performance of the \ac{backcom} system.
     \item Notably, our generalized analytical expressions include the existing results for Rayleigh fading channels as a special case when $m=1$.

 \end{enumerate}
%%%% paper structure
The structure of the paper is as follows: Section~\ref{system_model} introduces the system and channel models. Section~\ref{performance_analysis} details the performance evaluation for the considered system under various tag selection protocols. Numerical results and their analysis are provided in Section~\ref{numerical_results}. Finally, the conclusions are summarized in Section~\ref{conclusion}.
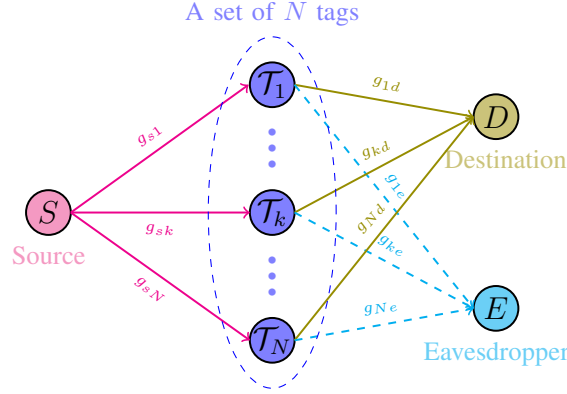
\begin{figure}[t!]
% \resizebox{8cm}{4.5cm}{%
	\begin{center}
		\begin{tikzpicture}[scale=0.85]
       % \draw[dotted] (0,0 )grid (8,5);
        %%% bullet T_1 and T_k
        \node [blue!50] at (4,3.25) {\tiny $\bullet$};
        \node [blue!50] at (4,3.0) {\tiny $\bullet$};
        \node [blue!50] at (4,2.75) {\tiny $\bullet$};
         %%% bullet T_k and T_n
        \node [blue!50] at (4,1.25) {\tiny $\bullet$};
        \node [blue!50] at (4,1.0) {\tiny $\bullet$};
        \node [blue!50] at (4,0.75) {\tiny $\bullet$};
		%%%Source S
		\draw[line width=0.75pt, fill=magenta!50](0.5,2.0) circle(0.35);
		\node [] at (0.5,2.0) {$S$};
        \node [below,magenta!50] at (0.5,1.65) {\small Source};
		%%%Destination D
		\draw[line width=0.75pt,fill=olive!50](7.5,3.5) circle(0.35);
		\node [] at (7.5,3.5) { $D$};
         \node [below,olive!50] at (7.65,3.15) {\small Destination};
        %%% Eavesdropper E
		\draw[line width=0.75pt,fill=cyan!50](7.5,0.5) circle(0.35);
		\node [] at (7.5,0.5) { $E$};
        \node [below,cyan!50] at (7.5,0.15) {\small Eavesdropper};
		%%%% Tag T1
		\draw[line width=0.75pt,fill=blue!50](4,4.0) circle(0.35);
		\node [] at (4,4.0) {$\mathcal{T}_1$};
        %%% S-T_1
		\draw[magenta, line width=0.75pt][->] (0.85,2)--(3.63,4)node [midway, above, sloped] (TextNode) {\tiny $g_{s1}$};
        %%% T_1-D
		\draw[olive, line width=0.75pt][->] (4.35,4)--(7.15,3.5) node [midway, above, sloped] (TextNode) {\tiny $g_{1d}$};
         %%% T_1-E
		\draw[dashed, cyan, line width=0.75pt][->] (4.35,4)--(7.15,0.5) node [midway, above, sloped] (TextNode) {\tiny $g_{1e}$};
		% \draw[dotted, line width=0.75pt][] (4,2.1)--(4,1.4);
		%%%% Tag T_K
		\draw[line width=0.75pt,fill=blue!50](4,2) circle(0.35);
		\node [] at (4,2.0) { $\mathcal{T}_k$};
        %%% S-T_k
		\draw[magenta,line width=0.75pt][->] (0.85,2.0)--(3.65,2.0)node [midway, below, sloped] (TextNode) {\tiny $g_{sk}$};
         %%% T_k-D
		\draw[olive, line width=0.75pt][->] (4.35,2)--(7.15,3.5) node [midway, above, sloped] (TextNode) {\tiny $g_{kd}$};
         %%% T_k-E
		\draw[dashed, cyan, line width=0.75pt][->] (4.35,2)--(7.15,0.5) node [midway, above, sloped] (TextNode) {\tiny $g_{ke}$};
		%%%% Tag T_N
		\draw[line width=0.75pt,fill=blue!50](4,0) circle(0.35);
		\node [] at (4,0) {$\mathcal{T}_N$};
         %%% S-T_N
		\draw[magenta,line width=0.75pt][->] (0.85,2)--(3.65,0)node [midway, below, sloped] (TextNode) {\tiny $g_{sN}$};
        %%% T_N-D
		\draw[olive, line width=0.75pt][->] (4.35,0)--(7.15,3.5) node [midway, above, sloped] (TextNode) {\tiny $g_{Nd}$};
         %%% T_N-E
		\draw[dashed, cyan, line width=0.75pt][->] (4.35,0)--(7.15,0.5) node [midway, above, sloped] (TextNode) {\tiny $g_{Ne}$};
		% %%%%%%%%%%%
		\draw [dashed, blue] (4,2.) ellipse (1cm and 2.75cm);
        \node [above,blue!50 ] at (4,4.75) {\small A set of $N$ tags};
		\end{tikzpicture}
	\end{center}
	\caption{System model illustrating energy-harvesting backscatter communications with multiple tags, a single destination, and an eavesdropper}
	 \label{sysmodel}
\end{figure}

\section{System and Channel Model}
\label{system_model}
%%% Corrected
In this study, we consider a \ac{backcom} system as shown in Fig.~\ref{sysmodel}. It has a source node $S$, $N$ tags $\mathcal{T}_k$ that have energy harvesting units, where $k \in \{1,2,...,N\}$, a destination node $D$, and an eavesdropper node $E$. Each node in the system is assumed to operate with a single antenna configuration. Each tag contains distinct information that needs to be conveyed to $D$, ensuring each tag's data is unique. Therefore, $S$ selects a single tag per transmission to relay its specific information. Let $\mathcal{X} \in \{D,E\}$, with the link between $S$ and $\mathcal{T}_k$ denoted as $S-\mathcal{T}_k$ and the link between $\mathcal{T}_k$ and $\mathcal{X}$ represented as $\mathcal{T}_k- \mathcal{X}$. The complex channel coefficient for $S-\mathcal{T}_k$ link can be expressed as $h_{sk} = \sqrt{d'^{\,-u_{sk}}_{sk}} \, g_{sk} \, e^{-j \tilde{\Theta}_{sk}}$, where $d'_{sk}$, $u_{sk}$, $g_{sk}$ and $\tilde{\Theta}_{sk}$ corresponds to the distance, path loss exponent, fading gain and phase, respectively. Similarly, the $\mathcal{T}_k-\mathcal{X}$ link's channel coefficient can be expressed as $h_{kx} = \sqrt{d'^{\,-u_{kx}}_{kx}} \, g_{kx} \, e^{-j \tilde{\Theta}_{kx}}$, with $d'_{kx}$, $u_{kx}$, $g_{kx}$ and $\tilde{\Theta}_{kx}$ representing the distance, path loss exponent, fading gain, and phase.  We assume that the channel gains $g_{sk}$ and $g_{kx}$ for each $k$ are \ac{inid}, following a Nakagami-$m$ fading model. The phases $\tilde{\Theta}_{sk}$ and $\tilde{\Theta}_{kx}$
are independently uniformly distributed over the interval $-\pi$ to $\pi$. The \ac{pdf} for each $g_{sk}^2$ and $g_{kx}^2$ can be given as follows \cite[(2)]{dixit2014performance}:
\begin{align}
    \label{pdfgsk}
f_{g_{sk}^2}\left(t\right)&=\frac{\tilde{\lambda}_{sk}^{m_{sk}}t^{m_{sk}-1}}{\Gamma\left(m_{sk}\right)}\exp\left(-\tilde{\lambda}_{sk}t\right),
\end{align}
\begin{align}
    \label{pdfgkx}
f_{g_{kx}^2}\left(t\right)=\frac{\tilde{\lambda}_{kx}^{m_{kx}}t^{m_{kx}-1}}{\Gamma\left(m_{kx}\right)}\exp\left(-\tilde{\lambda}_{kx}t\right),
\end{align}
where $\Gamma(\cdot)$ denotes the gamma function, $\tilde{\lambda}_{sk}=m_{sk}/\Omega_{sk}$, $\tilde{\lambda}_{kx}=m_{kx}/\Omega_{kx}$,  $\Omega_{sk}$ and $\Omega_{kx}$ represent average value of $g^2_{sk}$ and $g^2_{kx}$, respectively, and $m_{sk}$ and $m_{kx}$ represent fading parameter for $g^2_{sk}$ and $g^2_{kx}$, respectively \cite{SimonBookDC}. By applying the formulas \cite[(8.351.2),(8.352.1)]{gradshteyn2007table} into \cite[(3)]{dixit2014performance},  the corresponding \ac{cdf} expression for each $g_{sk}^2$ and $g_{kx}^2$ can be given as 
\begin{align}
    \label{cdfgsk}
F_{g_{sk}^2}\left(t\right)&=\mathbb{P}\left( g_{sk}^2 < t\right)
=\frac{\gamma\left(m_{sk},\tilde\lambda_{sk}t\right)}{\Gamma\left(m_{sk}\right)}
% \nonumber\\&
=1-
\sum_{j=0}^{m_{sk}-1}\frac{\exp\left(-\tilde{\lambda}_{sk}t\right)}{\left(\tilde{\lambda}_{sk}t\right)^{-j}\Gamma(j+1)},
\end{align}
and
\begin{align}
    \label{cdfgkx}
F_{g_{kx}^2}\left(t\right)&=\mathbb{P}\left( g_{kx}^2 < t\right)
=\frac{\gamma\left(m_{kx},\tilde\lambda_{kx}t\right)}{\Gamma\left(m_{kx}\right)}
% \nonumber\\&
=1-
\sum_{j=0}^{m_{kx}-1}\frac{\exp\left(-\tilde{\lambda}_{kx}t\right)}{\left(\tilde{\lambda}_{kx}t\right)^{-j}\Gamma(j+1)},
\end{align}
respectively, where $\mathbb{P}\left[\cdot\right]$ refers to the probability of an event occurring and $\gamma(\cdot,\cdot)$ is the lower incomplete gamma function \cite[8.350.1]{gradshteyn2007table}. 

Let $m_s$ represent the unit energy message symbol sent by the source node $S$, and the received signal at $\mathcal{T}_k$ is then given by
 \begin{align}
 \label{ykt}
     y_k=\sqrt \mathcal{P}\ h_{sk}\,m_s,
 \end{align}
 where $\mathcal{P}$ is the transmission power of $S$. The noise in the received signal at $\mathcal{T}_k$ is assumed negligible, as it is substantially lower than the RF signal strength \cite{wang2016ambient}. The signal received at $\mathcal{T}_k$ is further partition into two parts: $\sqrt{{1-\delta}_n}y_k(t)$ for \ac{eh} and $\sqrt{\delta_n}y_k(t)$ for \ac{backcom}, where $\delta_n$ represents the dynamic optimal reflection coefficient of $\mathcal{T}_k$. The input power for the energy harvesting unit at $\mathcal{T}_k$ is given by $\mathcal{P}_k^i=\left(1-\delta_k\right)\mathcal{P}d'^{\,-u_{sk}}_{sk}g_{sk}^2$. For the non-linear \ac{eh} model, the harvested power at $\mathcal{T}_k$
 is represented as \cite{ye2020outage}
\begin{align}
\label{pko}
\mathcal{P}_k^o&=\frac{\mathcal{P}_{\mathcal{T}_k}^{max}\left(1-\exp\left(-\xi_1\mathcal{P}_k^i+\xi_1\xi_0\right)\right)}{1+\exp\left(-\xi_1\mathcal{P}_k^i+\xi_1\xi_2\right)},
\end{align}
where $\mathcal{P}_{\mathcal{T}_k}^{max}$ indicates the saturation power level when $\mathcal{P}_k^i$ is considerably high and $\xi_0$, $\xi_1$, and $\xi_2$ represent the sensitivity threshold, resistance, capacitance, respectively.
%%%%%%
 $\mathcal{P}_{\mathcal{T}_k}^c$ refers to the power consumed by the tag's circuit. Due to the energy causality, $k$-th tag $\mathcal{T}_k$ can only reflect the source's signal to $D$ if $\mathcal{P}_k^o\geq\mathcal{P}_{\mathcal{T}_k}^c$. Consequently, to optimize the power of the backscatter signal, $\mathcal{P}_k^o=\mathcal{P}_{\mathcal{T}_k}^c$ must be satisfied. Considering $0\leq\beta_k^\ast\leq1$, the optimal dynamic reflection coefficient \cite{ye2020outage} can be formulated as
\begin{align}
\label{betakast}
    \beta_k^\ast&=\max\left(1-\frac{\phi}{\mathcal{P}{d'^{\,-u_{sk}}_{sk}}g_{sk}^2},0\right),
\end{align}
where $\phi=\xi_1^{-1}\ln (\phi_1/\phi_2)$,   $\phi_1=\mathcal{P}_{\mathcal{T}_k}^{max}\exp(\xi_1\xi_2)+\mathcal{P}_{\mathcal{T}_k}^c \exp(\xi_1\xi_0)$, $\phi_2=\mathcal{P}_{\mathcal{T}_k}^{max}-\mathcal{P}_{\mathcal{T}_k}^c$. $ \beta_k^\ast$ in As seen from (\ref{betakast}), when $\beta_k^\ast>0$, $\mathcal{T}_k$  has sufficient energy to reflect the signal, however, if $\beta_k^\ast=0$, $\mathcal{T}_k$ is unable to perform the reflection. Using 
 (\ref{ykt}) and (\ref{betakast}), the backscatter signal at $\mathcal{T}_k$ is given by 
\begin{align}
    \hat{y}_k&=\ \sqrt{\zeta\beta_k^\ast}y_k c_k=\sqrt{\zeta\beta_k^\ast\mathcal{P}}h_{sk}m_s c_k
\end{align}
where $c_k$ is the transmitted signal from $\mathcal{T}_k$ with unit energy, and $\zeta$ is the tag scattering coefficient \cite{wang2016ambient}. The signal received at $\mathcal{X}$ can be formulated as 
\begin{align}
\label{yx}
y_x&=\sqrt{\zeta\beta_k^\ast\mathcal{P}}h_{sk}h_{kx}m_s c_k+w_x,
\end{align}
where $w_x$ is the \ac{awgn} at $\mathcal{X}$, having a zero mean and a variance of $\sigma^2$. Thus, the instantaneous \ac{snr} at $\mathcal{X}$ can be given as 
\begin{align}
\label{snrxk}
    \gamma_x^k&=\zeta\beta_k^\ast {d'^{\,-u_{sk}}_{sk}}{d'^{\,-u_{kx}}_{kx}}g_{sk}^2\,g_{kx}^2\frac{\Gamma_t}{\Gamma_p},
\end{align}
where $\Gamma_t=\frac{\mathcal{P}}{\sigma^2}$ represents the average transmit SNR and $\Gamma_p$ denotes the performance gap due to the applied modulation scheme.

\section{Performance Analysis with Tag Selection}
\label{performance_analysis}
This section provides both exact and asymptotic analyses of \ac{sop} and \ac{ip} for various tag selection protocols.
%%% 
% \subsection{Exact \ac{sop} Analysis}
\subsection{Exact \texorpdfstring{\acs{sop}}{SOP} Analysis}
The \ac{sop} of the considered \ac{backcom} system model is defined as the probability that the secrecy rate $\mathcal{C}_{sec}\left(\gamma_d^k,\gamma_e^k\right)$ falls below a specified threshold rate $\mathcal{R}$, i.e. \cite{khan2024secrecy}
\begin{align}
\label{sopgen}
\widetilde{P}\left(\mathcal{R}\right)=\mathbb{P}\left[\mathcal{C}_{sec}\left(\gamma_d^k,\gamma_e^k\right)<\mathcal{R}\right],
\end{align}
where 
% $\mathbb{P}\left[\cdot\right]$ refers to the probability of an event occurring, 
% and 
$\mathcal{C}_{sec}\left(\gamma_d^k,\gamma_e^k\right)$ is expressed as \cite[(8)]{khan2024secrecy}
\begin{align}
\label{csec}
\mathcal{C}_{sec}\left(\gamma_d^k,\gamma_e^k\right)&=\max\left(\log_2\left(\frac{1+\gamma_d^k}{1+\gamma_e^k}\right), 0\right).
\end{align}
The SOP for the $k$-th tag can be categorized into two cases for each selection scheme: Case 1, where the harvested power at $\mathcal{T}_k$  is insufficient to power its circuit, indicated by $\beta_k^\ast=0$; Case 2, where the harvested power meets $\mathcal{T}_k$'s minimum power requirements, but the resulting secrecy capacity is below the threshold $\mathcal{R}$, represented by $\beta_k^\ast>0$ and $\mathcal{C}_{sec}\left(\gamma_d^k,\gamma_e^k\right)<\mathcal{R}$.
%%%%=== Exact Analysis of SOTS Protocol
% \subsubsection{\ac{sots} Protocol}
\subsubsection{\texorpdfstring{\ac{sots}}{SOTS} Protocol}
\label{exactsots}
In the SOTS protocol, the tag selected for communication is the one with the highest power gain among all $h_{kd}$ links, defined as:
\begin{align}
\label{kastsots}
    k^\ast &= \arg \max_{1 \le k \le N} \left\{ g_{kd}^2 \right\}.
\end{align}
Here, since $g_{kd}^2$ variables are \ac{iid}, the \ac{cdf} of $g_{k^\ast d}^2$ can be written using order statistics as \cite[(4)]{dixit2014performance}
\begin{align}
\label{cdfsots1}
    F_{g_{k^\ast d}^2}\left(t\right)&=\left(F_{g_{kd}^2}(t)\right)^N.
\end{align}
Using (\ref{cdfgkx}) into (\ref{cdfsots1}) and simplifying the result using multinomial theorem \cite[(24.1.2.B)]{abramowitz1968handbook}, the \ac{cdf} of $g_{k^\ast d}^2$ is expressed as
\begin{align}
\label{cdfgkastd2}
&F_{g_{k^\ast d}^2}\left(t\right)= \left(1-
\sum_{j=0}^{m_{kd}-1}\frac{\exp\left(-\tilde{\lambda}_{kd}t\right)}{\left(\tilde{\lambda}_{kd}t\right)^{-j}\Gamma(j+1)}\right)^N
\nonumber\\&
=\sum_{\substack{n_1,n_2,\ldots,n_{m_{kd}+1}\geq 0 \\ n_1+n_2+\ldots+n_{m_{kd}+1}=N}}\delta\left(N, \theta_1,\theta_2,\tilde\lambda_{kd}\right)t^{\theta_2} 
% \nonumber \\ &\times 
\exp\left(-\tilde\lambda_{kd}\theta_1 t\right),
\end{align}
where $\theta_1=n_2+n_3+\ldots+n_{m_{kd}+1}$, $\theta_2=n_3+2n_4+\ldots$ $+(m_{kd}-1)n_{m_{kd}+1}$, and 
\begin{align*}
 & \delta\left(N, \theta_1,\theta_2,\tilde\lambda_{kd}\right)= \frac{\left(-1\right)^{\theta_1} \tilde\lambda_{kd}^{\theta_2}\Gamma(N+1)}{\left(\prod_{i=1}^{m_{kd}+1}\Gamma(n_i+1)\right)\left(\prod_{i=1}^{m_{kd}}\left(\Gamma(i)\right)^{n_{i+1}}\right)}.
\end{align*}
Applying results from (\ref{sopgen}) and (\ref{kastsots}), the SOP for the SOTS protocol is formulated as \cite{khan2024secrecy}
\begin{align}
&\widetilde{P}_{out}= \underbrace{\mathbb{P}\left(\beta_k^\ast=0\right)}_{=\widetilde{P}_1} + \underbrace{\mathbb{P}\left(\beta_k^\ast>0, \ \mathcal C_{sec}\left(\gamma_d^{k^\ast}, \gamma_e^k\right)<\mathcal{R}\right)}_{=\widetilde{P}_2}.
\end{align}
Using (\ref{cdfgsk}) and (\ref{betakast}), $\widetilde{P}_1 $ can be represented as
\begin{align}
\label{P1sots}
\widetilde{P}_1& = \mathbb{P}\left(\max\left(1 - \frac{\phi}{\mathcal{P}{d'^{\,-u_{sk}}_{sk}}|g_{sk}|^2}, 0\right) = 0\right)
% \nonumber \\&
=\mathbb{P}\left( g_{sk}^2 < \frac{\phi}{\mathcal{P} d_{{sk}}'^{-u_{{sk}}}} \right)
 \nonumber \\&
=F_{g_{sk}^2}\left(\frac{\phi}{\mathcal{P} d_{{sk}}'^{-u_{{sk}}}} \right)
% \nonumber \\&
=\frac{\gamma\left(m_{sk},\frac{\tilde\lambda_{sk}\phi}{\mathcal{P} d'^{-u_{sk}}_{sk}}\right)}{\Gamma\left(m_{sk}\right)}.
\end{align}
%%%%%%
Using (\ref{betakast}) and (\ref{csec}), $\widetilde{P}_2 $ can be represented as
\begin{align}
\label{P2sotsa}
\widetilde{P}_2 &= {\mathbb{P}\left(\beta_k^\ast>0, \ \mathcal C_{sec}\left(\gamma_d^{k^\ast}, \gamma_e^k\right)<\mathcal{R}\right)}
% \nonumber\\
%  &
 =\mathbb{P}\left( g_{sk}^2 <\frac{\phi}{\mathcal{P} d_{{sk}}'^{-u_{{sk}}}} , 
 \frac{1 + \beta_k^\ast \Gamma_t \eta_1\ g_{sk}^2 g_{k^*d}^2}{1 + \beta_k^\ast \Gamma_t \eta_2 g_{sk}^2 g_{ke}^2} < \tau 
 \right),
\end{align}
where $\tau=2^\mathcal{R}$, $\eta_1=\frac{\zeta{d'^{\,-u_{sk}}_{sk}}{d'^{\,-u_{kd}}_{kd}}}{\Gamma_p}$, and $\eta_2=\frac{\zeta{d'^{\,-u_{sk}}_{sk}}{d'^{\,-u_{ke}}_{ke}}}{\Gamma_p}$. In deriving $\widetilde{P}_2$, the transformations of the random variables $W_1=g_{sk}^2$, $W_2=g_{k^\ast d}^2$, and $W_3=g_{ke}^2$ are considered. We can reformulate $\widetilde{P}_2$ as
\begin{align}
\label{P2sostb}
\widetilde{P}_2&=\mathbb{P}\left(W_1>\frac{\phi}{\mathcal{P}{d'^{\,-u_{sk}}_{sk}}},W_2<\frac{\tau-1}{\beta_n^\ast\eta_1\Gamma_t w_1}+\frac{\tau\eta_2 w_3}{\mathrm{\eta_1}}\right)\nonumber 
\\&=\int_{0}^{\infty}\int_{\frac{\phi}{\mathcal{P}{d'^{\,-u_{sk}}_{sk}}}}^{\infty}F_{W_2}\left(\frac{\tau-1}{\beta_n^\ast\eta_1 \Gamma_t w _1}+\frac{\tau\eta_2 w_3}{\eta_1}\right) f_{W_1}\left(w_1\right)
% \nonumber\\&\times 
f_{W_3}\left(w_3\right)dw_1 dw_3. 
\end{align}
By substituting $F_{W_2}(\cdot)$, $f_{W_1}(\cdot)$ and $f_{W_3}(\cdot)$ into (\ref{P2sostb}) and  evaluating the resulting integrals, a closed-form expression for $\widetilde{P}_2$ can be obtained, as 
% shown in (\ref{P2sotsFinal}) on the next page, 

% \begin{figure*}[t!]
% \normalsize
% \setcounter{equation}{19}
% For SOTS:
\begin{align}\label{P2sotsFinal}
\widetilde{P}_2&= \left( \frac{\tilde{\lambda}_{sk}^{m_{sk}} \tilde{\lambda}_{ke}^{m_{ke}}}{\Gamma(m_{sk}) \Gamma(m_{ke})} \right) 
\sum_{\substack{n_1, n_2, \ldots, \\ n_{m_{kd}+1} \geq 0 \\ n_1 + n_2 + \cdots + n_{m_{kd}+1} = N}} 
\delta(N, \theta_1, \theta_2, \tilde{\lambda}_{kd})  \sum_{q=0}^{\theta_2} \binom{\theta_2}{q} \left( \frac{\tau \eta_2}{\eta_1} \right)^q  
\exp\left( - \frac{\tilde{\lambda}_{sk}\phi}{\mathcal{P} d'^{-u_{sk}}_{sk}} \right) 
\nonumber\\
& \times
\left( \frac{\tau - 1}{\eta_1 \Gamma_t} \right)^{\theta_2 - q}
% \nonumber \\
% & \times 
\sum_{p=0}^{m_{sk} - 1} \binom{m_{sk} - 1}{p} 
\left( \frac{\phi}{\mathcal{P} d'^{-u_{sk}}_{sk}} \right)^{m_{sk} - 1 - p} 
2 \left( \frac{\tilde{\lambda}_{kd} \theta_1 (\tau - 1)}{\eta_1 \Gamma_t \tilde{\lambda}_{sk}} 
\right)^{\frac{p + q - \theta_2 + 1}{2}}
\nonumber\\
& \times
K_{p + q - \theta_2 + 1} \left( \sqrt{\frac{4 \theta_1 \tilde{\lambda}_{sk} \tilde{\lambda}_{kd} 
(\tau - 1)}{\eta_1 \Gamma_t}} \right) %\right) 
% \nonumber\\
% & \times
\left(\tilde{\lambda}_{ke} + \frac{\tilde{\lambda}_{kd} \theta_1 \tau \eta_2}{\eta_1} 
\right)^{-m_{ke} - q} \Gamma(m_{ke} + q),
% \end{aligned}
\end{align}
% \hrulefill
% \end{figure*}
where $K_u\left(\cdot\right)$ represents the modified Bessel function of the second kind of $u$-th order \cite{gradshteyn2007table}. A detail solution of $\widetilde{P}_2$ for the SOTS protocol is given in Appendix~\ref{sotsappendix}. The exact \ac{sop} for the \ac{sots} protocol is provided in Table~\ref{exactsoptable} on page 12.

%%%%=== Exact Analysis of METS Protocol
\subsubsection{METS Protocol}
In this protocol, the tag is chosen based on identifying the weakest 
$\mathcal{T}_k-E$ link. This selection method for minimizing eavesdropping can be represented as
\begin{equation}
\label{kastmets}
    k^\ast = \arg \min_{1 \le k \le N} \left\{ g_{ke}^2 \right\}
\end{equation}
Here, since $g_{ke}^2$ variables are \ac{iid}, the \ac{cdf} of $g_{k^\ast e}^2$ can be written using binomial expansion as
\begin{align}
    \label{cdfmets1}
    F_{g_{k^\ast e}^2}(t)&=1-\left(1-\mathbb{P}\left( g_{ke}^2 < t\right)\right)^N
    % \nonumber\\&
    =
    1-\left(1-F_{g_{k e}^2}(t)\right)^N
    \nonumber\\&=\sum_{l=1}^{N}\binom{N}{l}(-1)^{l+1}\left(F_{g_{k e}^2}(t)\right)^l
\end{align}
Using (\ref{cdfgkx}) into (\ref{cdfmets1}) and simplifying the result using multinomial theorem \cite[(24.1.2.B)]{abramowitz1968handbook}, the \ac{cdf} of $g_{k^\ast e}^2$ is expressed as 
% (\ref{cdfgkastd}) on the next page, 
% \begin{figure*}[t!]
% \normalsize
% \setcounter{equation}{22}
\begin{align}
\label{cdfgkastd1}
F_{g_{k^\ast e}^2}\left(t\right)&=\sum_{l=1}^{N}\binom{N}{l}(-1)^{l+1} 
% \nonumber\\&\times 
\left(1-
\sum_{j=0}^{m_{ke}-1}\frac{\exp\left(-\tilde{\lambda}_{kd}t\right)}{\left(\tilde{\lambda}_{ke}t\right)^{-j}\Gamma(j+1)}\right)^l
\nonumber\\&
=\sum_{l=1}^{N}\binom{N}{l}(-1)^{l+1} \sum_{\substack{n_1,n_2,\ldots,n_{m_{ke}+1}\geq 0 \\ n_1+n_2+\ldots+n_{m_{ke}+1}=l}}\delta_2\left(l, \theta_3,\theta_4,\tilde\lambda_{ke}\right)
% \nonumber\\& \times 
t^{\theta_4} 
\exp\left(-\tilde\lambda_{ke}\theta_3 t\right). 
\end{align}
% \hrulefill
% \end{figure*}
where $\theta_3=n_2+n_3+\ldots+n_{m_{ke}+1}$, $\theta_4=n_3+2n_4+\ldots+(m_{ke}-1)n_{m_{ke}+1}$, and 
\begin{align*}
 & \delta_2\left(l, \theta_3,\theta_4,\tilde\lambda_{ke}\right)= \frac{\left(-1\right)^{\theta_3} \tilde\lambda_{ke}^{\theta_4}\Gamma(l+1)}{\left(\prod_{i=1}^{m_{ke}+1}\Gamma(n_i+1)\right)\left(\prod_{i=1}^{m_{ke}}\left(\Gamma(i)\right)^{n_{i+1}}\right)}.
\end{align*}
The \ac{pdf} of $g_{k^\ast e}^2$ can be obtained by differentiating $F_{g_{k^\ast e}^2}\left(t\right)$ with respect to $t$ which is given in (\ref{pdfgkastd}) on page 5.
% \begin{figure*}[t!]
% \normalsize
% \setcounter{equation}{23}
\begin{align}
\label{pdfgkastd}
f_{g_{k^\ast e}^2}\left(t\right)&=\sum_{l=1}^{N}\binom{N}{l}(-1)^{l+1} 
\sum_{\substack{n_1,n_2,\ldots,n_{m_{ke}+1}\geq 0 \\ n_1+n_2+\ldots+n_{m_{ke}+1}=l}}\delta_2\left(l, \theta_3,\theta_4,\tilde\lambda_{ke}\right) 
 \nonumber \\ &\times 
\left(\theta_4 t^{\theta_4-1} \exp\left(-\tilde\lambda_{ke}\theta_3 t\right)-\tilde\lambda_{ke}\theta_3 t^{\theta_4} \exp\left(-\tilde\lambda_{ke}\theta_3 t\right) \right).
\end{align}
% \hrulefill
% \end{figure*}
Applying results from (\ref{sopgen}) and (\ref{kastmets}), the SOP for the METS protocol is formulated as \cite{khan2024secrecy}
\begin{align}
&\widetilde{P}_{out} = \underbrace{\mathbb{P}\left(\beta_k^\ast=0\right)}_{=\widetilde{P}_1} + \underbrace{\mathbb{P}\left(\beta_k^\ast>0, \ \mathcal C_{sec}\left(\gamma_d^{k}, \gamma_e^{k^\ast}\right)<\mathcal{R}\right)}_{=\widetilde{P}_2}.
\end{align}
Using a method akin to the SOTS, the expression for $\widetilde{P}_1$ is provided in (\ref{P1sots}). Using (\ref{sopgen}) and (\ref{kastmets}), $\widetilde{P}_2$ can be formulated as
\begin{align}
\widetilde{P}_2& = {\mathbb{P}\left(\beta_k^\ast>0, \ \mathcal C_{sec}\left(\gamma_d^{k^\ast}, \gamma_e^k\right)<\mathcal{R}\right)}
% \nonumber
% \nonumber\\&
=\mathbb{P}\left( g_{sk}^2 < \frac{\phi}{\mathcal{P} d_{sk}'^{-u_{sk}}}, \frac{1 + \beta_k^\ast \Gamma_t \eta_1\ g_{sk} ^2 g_{kd}^2}{1 + \beta_k^\ast \Gamma_t \eta_2 g_{sk}^2 g_{k^*e}^2} < \tau \right).
% \nonumber
\end{align}
In deriving $\widetilde{P}_2$, the transformations of the random variables $W_1=g_{sk}^2$, $W_2=g_{k^\ast d}^2$, and $W_3=g_{ke}^2$ are considered. We can reformulate $\widetilde{P}_2$ as
\begin{align}
\label{P2metsb}
\widetilde{P}_2&=\mathbb{P}\left(W_1>\frac{\phi}{\mathcal{P}{d'^{\,-u_{sk}}_{sk}}},W_2<\frac{\tau-1}{\beta_n^\ast\eta_1\Gamma_t w_1}+\frac{\tau\eta_2 w_3}{\mathrm{\eta_1}}\right)\nonumber 
\\&=\int_{0}^{\infty}\int_{\frac{\phi}{\mathcal{P}{d'^{\,-u_{sk}}_{sk}}}}^{\infty}F_{W_2}\left(\frac{\tau-1}{\beta_n^\ast\eta_1 \Gamma_t w _1}+\frac{\tau\eta_2 w_3}{\eta_1}\right) f_{W_1}\left(w_1\right)
% \nonumber\\&\times 
f_{W_3}\left(w_3\right)dw_1 dw_3. 
\end{align}
By substituting $F_{W_2}(\cdot)$, $f_{W_1}(\cdot)$ and $f_{W_3}(\cdot)$ into (\ref{P2metsb}) and  evaluating the resulting integrals, a closed-form expression for $\widetilde{P}_2$ can be obtained, as
% shown in (\ref{P2metsFinal}) on the next page. 
% \begin{figure*}[t!]
% \normalsize
% \setcounter{equation}{27}
% For METS:
\begin{align}
\label{P2metsFinal}
&\widetilde{P}_2=
1 - \frac{\gamma\left(m_{sk}, \frac{\tilde\lambda_{sk} \phi}{\mathcal{P} \tilde\lambda_{sk}^{-v_{sn}}}\right)}{\Gamma(m_{sn})} - \sum_{j=0}^{m_{kd}-1} \frac{\tilde\lambda_{kd}^j}{\Gamma(j+1)} \sum_{q=0}^j \binom{j}{q} \left(\frac{\tau \eta_2}{\eta_1}\right)^q   \left( \frac{\tau-1}{\eta_1 \Gamma_t} \right)^{j-q} \frac{\tilde\lambda_{sk}^{m_{sk}}}{\Gamma(m_{sk})} \exp\left( -\frac{\tilde{\lambda}_{sk} \phi}{\mathcal{P} d'^{-u_{sk}}_{sk}} \right) \nonumber \\
& \times \sum_{p=0}^{m_{{sk}} - 1} \binom{m_{{sk}} - 1}{p} 
\left( \frac{\phi}{\mathcal{P} d'^{-u_{sk}}_{sk}} \right)^{m_{sk} - 1 - p} 2 \left( \frac{\tilde\lambda_{kd} (\tau-1)}{\tilde\lambda_{sk} \eta_1 \Gamma_t} \right)^{\frac{p+q-j+1}{2}} K_{p+q-j+1} \left( \sqrt{\frac{4 \tilde\lambda_{sk} \tilde\lambda_{kd} (\tau-1)}{\eta_1 \Gamma_t}} \right) 
\nonumber\\
& \times
\sum_{l=1}^{N}\binom{N}{l}(-1)^{l+1} 
% \nonumber\\
% & \times
\sum_{\substack{n_1,n_2,\ldots,n_{m_{ke}}\geq0 \\ n_1+n_2+\cdots+n_{m_{ke}}=l}} 
\delta\left(l, \theta_3, \theta_4, \tilde{\lambda}_{ke}\right)  %\left.
\left(\tilde\lambda_{ke} \theta_3 + \frac{\tilde\lambda_{kd} \tau \eta_2}{\eta_1} \right)^{-(\theta_4+q)} \Gamma(\theta_4+q+1)
\nonumber\\
& \times
\left( \frac{\tilde\lambda_{{ke}} \theta_3}{\tilde\lambda_{{ke}} \theta_3 + \frac{\tilde\lambda_{{kd}} \tau \eta_2}{\eta_1}} 
- \frac{\theta_4}{\theta_4 + q} \right).
\end{align}
% \hrulefill
% \end{figure*}
A detail solution of $\widetilde{P}_2$ for the METS protocol is given in Appendix~\ref{metsappendix}. 
% Finally, the exact \ac{sop} for the \ac{mets} protocol can be obtained as given in Table~\ref{exactsoptable}. 
Table~\ref{exactsoptable} presents the exact \ac{sop} for the \ac{mets} protocol on page 12.

%%% 
%%% Exact SOP for OTS Protocol
\subsubsection{OTS Protocol}
In the \ac{ots} method, the tag chosen is the one with the highest instantaneous secrecy capacity. The optimal tag is identified based on the condition 
\begin{align}
    k^\ast& = \arg \max_{1 \le k \le N} \mathcal{C}_{sec}\left(\gamma_d^k,\gamma_e^k\right).
\end{align}
% Since the channels are independent, the total SOP for this selection method is the product of the SOPs for each tag. 
Given the independence of all channels, the total \ac{sop} for this selection approach is derived as the product of the individual \ac{sop}s, ie, \cite{khan2024secrecy}
\begin{align}
\widetilde{P}_{out} = \left(\mathbb{P}\left(\beta_k^\ast=0\right)+ \mathbb{P}\left(\beta_k^\ast>0, \ \mathcal C_{sec}\left(\gamma_d^k, \gamma_e^k\right)<\mathcal{R}\right)\right)^N. 
\end{align}
Applying similar mathematical steps used for deriving the \ac{sop} of the \ac{mets} protocol, the \ac{sop} for the \ac{ots} protocol can be obtained as provided in Table~\ref{exactsoptable} on page 12.
%%% Exact SOP for RTS Protocol
\subsubsection{RTS Protocol}
In this selection approach, a tag is chosen at random, represented by the formulation
\begin{align}
    k^\ast &= \arg \max_{1 \le k \le N} \mathcal{C}_{sec}\left(\gamma_d^k,\gamma_e^k\right).
\end{align}
Applying similar mathematical steps used for deriving the \ac{sop} of the \ac{sots} protocol, the \ac{sop} for the \ac{rts} protocol can be obtained as given in Table~\ref{exactsoptable} on page 12.
% \vspace{-1cm}
%%%% Exact SOP Table-I
\begin{table*}[t!]
\centering
  % \label{ExactSOPtable1}
  \caption{Exact SOP Expressions}
    \label{exactsoptable}
% \resizebox{\columnwidth}{!}{%
\scalebox{0.85}{
\begin{tabular}{|l|l|}
\hline
Protocol &  \multicolumn{1}{c|}{SOP}\\ \hline
SOTS & $\begin{aligned}\widetilde{P}_{out}&= \frac{\gamma\left(m_{sk},\frac{\tilde\lambda_{sk}\phi}{\mathcal{P} d'^{-u_{sk}}_{sk}}\right)}{\Gamma\left(m_{sk}\right)} + \left( \frac{\tilde{\lambda}_{sk}^{m_{sk}} \tilde{\lambda}_{ke}^{m_{ke}}}{\Gamma(m_{sk}) \Gamma(m_{ke})} \right) 
\sum_{\substack{n_1, n_2, \ldots, n_{m_{kd}+1} \geq 0 \\ n_1 + n_2 + \cdots + n_{m_{kd}+1} = N}} 
\delta(N, \theta_1, \theta_2, \tilde{\lambda}_{kd})  \sum_{q=0}^{\theta_2} \binom{\theta_2}{q} \left( \frac{\tau \eta_2}{\eta_1} \right)^q  \nonumber \\
& \times \exp\left( - \frac{\tilde{\lambda}_{sk}\phi}{\mathcal{P} d'^{-u_{sk}}_{sk}} \right) 
\left( \frac{\tau - 1}{\eta_1 \Gamma_t} \right)^{\theta_2 - q} \sum_{p=0}^{m_{sk} - 1} \binom{m_{sk} - 1}{p} 
\left( \frac{\phi}{\mathcal{P} d'^{-u_{sk}}_{sk}} \right)^{m_{sk} - 1 - p} 
% \nonumber\\
% & \times \left(
2 \left( \frac{\tilde{\lambda}_{kd} \theta_1 (\tau - 1)}{\eta_1 \Gamma_t \tilde{\lambda}_{sk}} 
\right)^{\frac{p + q - \theta_2 + 1}{2}} 
\nonumber\\ & \times K_{p + q - \theta_2 + 1} \left( \sqrt{\frac{4 \theta_1 \tilde{\lambda}_{sk} \tilde{\lambda}_{kd} 
(\tau - 1)}{\eta_1 \Gamma_t}} \right) 
% \right) 
% \nonumber\\
% & \times
\left(\tilde{\lambda}_{ke} + \frac{\tilde{\lambda}_{kd} \theta_1 \tau \eta_2}{\eta_1} 
\right)^{-m_{ke} - q} \Gamma(m_{ke} + q).
\end{aligned}$ \\ \hline
METS & $\begin{aligned} \widetilde{P}_{out} &=
1 - \frac{\gamma\left(m_{sk}, \frac{\tilde\lambda_{sk} \phi}{\mathcal{P} \tilde\lambda_{sk}^{-v_{sn}}}\right)}{\Gamma(m_{sn})} - \sum_{j=0}^{m_{kd}-1} \frac{\tilde\lambda_{kd}^j}{\Gamma(j+1)} \sum_{q=0}^j \binom{j}{q} \left(\frac{\tau \eta_2}{\eta_1}\right)^q  \left( \frac{\tau-1}{\eta_1 \Gamma_t} \right)^{j-q} \frac{\tilde\lambda_{sk}^{m_{sk}}}{\Gamma(m_{sk})} \exp\left( -\frac{\tilde{\lambda}_{sk} \phi}{\mathcal{P} d'^{-u_{sk}}_{sk}} \right)\nonumber \\
& \times \sum_{p=0}^{m_{{sk}} - 1} \binom{m_{{sk}} - 1}{p} 
\left( \frac{\phi}{\mathcal{P} d'^{-u_{sk}}_{sk}} \right)^{m_{sk} - 1 - p} 2 \left( \frac{\tilde\lambda_{kd} (\tau-1)}{\tilde\lambda_{sk} \eta_1 \Gamma_t} \right)^{\frac{p+q-j+1}{2}} K_{p+q-j+1} \left( \sqrt{\frac{4 \tilde\lambda_{sk} \tilde\lambda_{kd} (\tau-1)}{\eta_1 \Gamma_t}} \right) \nonumber\\
& \times \sum_{l=1}^{N}\binom{N}{l}(-1)^{l+1} 
\sum_{\substack{n_1,n_2,\ldots,n_{m_{ke}}\geq0 \\ n_1+n_2+\cdots+n_{m_{ke}}=l}} 
\delta\left(l, \theta_3, \theta_4, \tilde{\lambda}_{ke}\right)
\left(\tilde\lambda_{ke} \theta_3 + \frac{\tilde\lambda_{kd} \tau \eta_2}{\eta_1} \right)^{-(\theta_4+q)} \Gamma(\theta_4+q+1) \\
& \times \left( \frac{\tilde\lambda_{ke} \theta_3}{\tilde\lambda_{ke} \theta_3 + \frac{\tilde\lambda_{kd} \tau \eta_2}{\eta_1}} 
- \frac{\theta_4}{\theta_4 + q} \right).
\end{aligned}$ \\ \hline
OTS &  $\begin{aligned}\widetilde{P}_{out} &= \left(1-\frac{\tilde{\lambda}_{sk}^{m_{sk}} \tilde{\lambda}_{ke}^{m_{ke}}}{\Gamma(m_{sk}) \Gamma(m_{ke})}  
\sum_{j=0}^{m_{kd}-1} \frac{\tilde\lambda_{kd}^j}{\Gamma\left(j+1\right)} 
\sum_{q=0}^{j} \binom{j}{q} \exp\left( - \frac{\tilde{\lambda}_{sk}\phi}{\mathcal{P} d'^{-u_{sk}}_{sk}} \right) \left(\frac{\tau-1}{\eta_1 \Gamma_t}\right)^{j-q} 
\sum_{p=0}^{m_{sk} - 1} \binom{m_{sk} - 1}{p} \right. \nonumber \\
&\left.\times \left( \frac{\phi}{\mathcal{P} d'^{-u_{sk}}_{sk}} \right)^{m_{sk} - 1 - p} 2 \left(\frac{\tilde{\lambda}_{kd} \left(\tau-1\right)}{\tilde{\lambda}_{sk} \eta_1 \Gamma_t}\right)^{\frac{p+q-j+1}{2}} 
K_{p+q-j+1}\left(\sqrt{\frac{4 \tilde{\lambda}_{sk} \tilde{\lambda}_{kd} \left(\tau-1\right)}{\eta_1 \Gamma_t}}\right) 
% \\
% & 
\left( \frac{\tau \eta_2}{\eta_1} \right)^q \Gamma\left(m_{ke}+q\right)
\right. \nonumber\\& \left. \times\left(\frac{1}{\tilde{\lambda}_{ke} + \frac{\tilde{\lambda}_{kd} \tau \eta_2}{\eta_1}}\right)^{m_{ke}+q}
\right)^{N}.
\end{aligned}$ \\ \hline
 RTS & $\begin{aligned} \widetilde{P}_{out}& = 1 - \frac{1}{N} \sum_{n=1}^{N} \left( \frac{\tilde\lambda_{sk}^{m_{sk}} \tilde\lambda_{ke}^{m_{ke}}}{\Gamma\left(m_{sk}\right) \Gamma\left(m_{ke}\right)} \right) \sum_{j=0}^{m_{kd}-1} \frac{\lambda_{kd}^j}{\Gamma\left(j+1\right)} 
\sum_{q=0}^{j} \binom{j}{q} \exp\left( - \frac{\tilde{\lambda}_{sk}\phi}{\mathcal{P} d'^{-u_{sk}}_{sk}} \right) \left(\frac{\tau-1}{\eta_1 \Gamma_t}\right)^{j-q} 
% \right.
\nonumber\\
& \times 
\sum_{p=0}^{m_{sk} - 1} \binom{m_{sk} - 1}{p} \left( \frac{\phi}{\mathcal{P} d'^{-u_{sk}}_{sk}} \right)^{m_{sk} - 1 - p}  2 \left( \frac{\tilde{\lambda}_{kd} \left(\tau-1\right)}{\tilde{\lambda}_{sk} \eta_1 \Gamma_t} \right)^{\frac{p+q-j+1}{2}}  K_{p+q-j+1} \left( \sqrt{\frac{4 \tilde{\lambda}_{sk} \tilde{\lambda}_{kd} \left(\tau-1\right)}{\eta_1 \Gamma_t}} \right)
\nonumber \\
& 
+\left( \frac{\tau \eta_2}{\eta_1} \right)^q \left( \frac{1}{\tilde{\lambda}_{ke} + \frac{\tilde{\lambda}_{kd} \tau \eta_2}{\eta_1}} \right)^{m_{ke}+q} \Gamma\left(m_{ke}+q\right).
% \right]
\end{aligned}$ \\ \hline
\end{tabular}
 }
\end{table*}

%%%%% Exact IP
% \subsection{Exact \ac{ip} Analysis}
\subsection{Exact \texorpdfstring{\ac{ip}}{IP} Analysis}
\label{IP}
In a \ac{backcom} network, the \ac{ip} represents the probability that the eavesdropper successfully intercepts. Mathematically, the \ac{ip} of any protocol can generally be written as \cite{khan2024secrecy}
\begin{align}
\label{ip1}
    \widetilde{P}_{ip}
    &= \underbrace{\mathbb{P}\left(\beta_k^\ast=0\right)}_{=\widetilde{P}_1} + \underbrace{\mathbb{P}\left(\beta_k^\ast>0, \ \gamma_d^{k}<\gamma_e^k\right)}_{=\widetilde{P}_3}.
\end{align}
 Following a similar mathematical approach as used for the \ac{sop} derivation, the \ac{ip} expressions for each selection method are derived and summarized in Table~\ref{exactiptable} on page 13.
%%%% Exact IP Table-II
\begin{table*}[t!]
\centering
  \caption{Intercept Probability Expressions}
   \label{exactiptable}
% \resizebox{\columnwidth}{!}{%
\scalebox{0.85}{
\begin{tabular}{|l|l|}
\hline
Protocol &  \multicolumn{1}{c|}{IP}\\ \hline
SOTS & $\begin{aligned}\widetilde{P}_{ip} &=  \frac{\gamma\left(m_{sk},\tilde\lambda_{sk}\frac{\phi}{\mathcal{P} d'^{-u_{sk}}_{sk}}\right)}{\Gamma\left(m_{sk}\right)}  + \frac{\tilde\lambda_{ke}^{m_{ke}} \Gamma\left(m_{sk}, \tilde\lambda_{sk}\frac{\phi}{\mathcal{P} d'^{-u_{sk}}_{sk}}\right)}{\Gamma\left(m_{sk}\right) \Gamma\left(m_{ke}\right)} \sum_{\substack{n_1,n_2,\ldots,n_{m_{kd}+1}\geq0 \\ n_1+n_2+\cdots+n_{m_{kd}+1}=N}} 
\delta\left(N, \theta_1, \theta_2, \tilde{\lambda}_{kd}\right) \left(\frac{\eta_2}{\eta_1}\right)^{\theta_2} \nonumber \\
& \times \left(\frac{1}{\tilde\lambda_{ke} + \frac{\tilde\lambda_{kd} \theta_1 \eta_2}{\eta_1}}\right)^{m_{ke}+\theta_2} \Gamma\left(m_{ke}+\theta_2\right).
\end{aligned}$ \\ \hline
METS & $\begin{aligned} 
\widetilde{P}_{ip} &= 1 - \sum_{j=0}^{m_{kd}-1} \left(\frac{\lambda_{kd}^j}{\Gamma(j+1)} \right)
\sum_{\substack{n_1,n_2,\ldots,n_{m_{ke}}\geq0 \\ n_1+n_2+\cdots+n_{m_{ke}}=l}} 
\delta\left(l, \theta_3, \theta_4, \tilde{\lambda}_{ke}\right) \left(\frac{\eta_2}{\eta_1}\right)^j \frac{\Gamma\left(m_{sk}, \tilde\lambda_{sk} \frac{\phi}{\mathcal{P} d'^{-u_{sk}}_{sk}}\right)}{\Gamma\left(m_{sk}\right)}  \\
& \times \left( \frac{1}{\lambda_{ke} \theta_3 + \frac{\lambda_{kd} \eta_2}{\eta_1}} \right)^{\theta_4 + j} \Bigg( \lambda_{ke} \theta_3 \left( \frac{1}{\lambda_{ke} \theta_3 + \frac{\lambda_{kd} \eta_2}{\eta_1}} \right) \Gamma\left(\theta_4 + j + 1\right) \quad - \theta_4 \Gamma\left(\theta_4 + j\right) \Bigg).
\end{aligned}$ \\ \hline
OTS & $\begin{aligned}
\widetilde{P}_{ip} &= \left( 1 - \left( \frac{\tilde\lambda_{ke}^{m_{ke}}}{\Gamma(m_{ke})} \right) \frac{\Gamma\left(m_{sk}, \tilde\lambda_{sk} \frac{\phi}{\mathcal{P} d'^{-u_{sk}}_{sk}}\right)}{\Gamma\left(m_{sk}\right)} \sum_{j=0}^{m_{kd}-1} \frac{\tilde\lambda_{kd}^j}{\Gamma(j+1)} \left(\frac{\eta_2}{\eta_1}\right)^j \right. \quad  \left( \frac{1}{\tilde\lambda_{ke} + \frac{\tilde\lambda_{kd} \eta_2}{\eta_1}} \right)^{m_{ke}+j} \nonumber \\ & \times \Gamma(m_{ke}+j) \Bigg)^{N}.
\end{aligned}$ \\ \hline
RTS & $\begin{aligned}
\widetilde{P}_{ip} &= \frac{1}{N} \sum_{n=1}^{N} \Bigg( 1 - \left( \frac{\tilde\lambda_{ke}^{m_{ke}}}{\Gamma(m_{ke})} \right) \frac{\Gamma\left(m_{sk}, \tilde\lambda_{sk} \frac{\phi}{\mathcal{P} d'^{-u_{sk}}_{sk}}\right)}{\Gamma\left(m_{sk}\right)} \quad \sum_{j=0}^{m_{kd}-1} \frac{\tilde\lambda_{kd}^j}{\Gamma(j+1)} \left(\frac{\eta_2}{\eta_1}\right)^j \left( \frac{1}{\tilde\lambda_{ke} + \frac{\tilde\lambda_{kd} \eta_2}{\eta_1}} \right)^{m_{ke}+j}  \nonumber \\ & \times \Gamma(m_{ke}+j) \Bigg).
\end{aligned}$ \\ \hline
\end{tabular}
 }
\end{table*}

\subsection{Asymptotic Analysis}
\label{ASYMPTOTIC AND COMPLEXITY ANALYSIS}
To explore the key factors influencing \ac{sop} and \ac{ip} more thoroughly, we conduct an asymptotic performance analysis under the assumption of very high transmission power, where $\Gamma_t$ or $\mathcal{P}$ approaches infinity. 
%\subsubsection{Asymptotic \ac{sop}} 
\subsubsection{Asymptotic \texorpdfstring{\ac{sop}}{SOP}}
The asymptotic \ac{sop} of any protocol can write as
\begin{align}
\label{asysop}
\widetilde{P}_{out}^{\infty}=\widetilde{P}_{1}^{\infty}+\widetilde{P}_{2}^{\infty},
\end{align}
where 
\begin{align}
\label{asysopP1}
\widetilde{P}_{1}^{\infty}=\lim_{\Gamma_t\rightarrow \infty}\widetilde{P}_{1}=\lim_{\mathcal{P}\rightarrow \infty}\frac{\gamma\left(m_{sk},\frac{\tilde\lambda_{sk}\phi}{\mathcal{P} d'^{-u_{sk}}_{sk}}\right)}{\Gamma\left(m_{sk}\right)}=0,
\end{align}
and
\begin{align}
\label{asysopP2}
&\widetilde{P}_{2}^{\infty}=\lim_{\Gamma_t\rightarrow \infty}\widetilde{P}_{2}
% \nonumber\\
% &
=\lim_{\Gamma_t\,\mbox{\small or}\,\mathcal{P}\rightarrow \infty}
\mathbb{P}\left(W_1>0,W_2<\frac{\tau\eta_2 w_3}{\mathrm{\eta_1}}\right)\nonumber 
\\&=\int_{0}^{\infty}\int_{0}^{\infty}F_{W_2}\left(\frac{\tau\eta_2 w_3}{\eta_1}\right) f_{W_1}\left(w_1\right)
% \nonumber\\&\times 
f_{W_3}\left(w_3\right)dw_1 dw_3.
\end{align}
% The asymptotic \ac{sop} for each protocol is obtained by solving (\ref{asysopP2}) using analogous mathematical techniques employed in deriving the exact \ac{sop} expressions. The resulting asymptotic \ac{sop} expressions for all selection protocols are summarized in Table~\ref{asysoptable}.
The asymptotic \ac{sop} for each protocol is derived from (\ref{asysopP2}) using similar techniques as the exact \ac{sop}. The asymptotic \ac{sop} expressions for all protocols are in Table~\ref{asysoptable} on page 14.
%%%% AsymptoticIP Table-III
\begin{table*}[t!]
\centering
  \caption{Asymptotic SOP Expressions}
    \label{asysoptable}
% \resizebox{\columnwidth}{!}{%
\scalebox{0.85}{
\begin{tabular}{|l|l|}
\hline
Protocol &  \multicolumn{1}{c|}{Asymptotic SOP}\\ \hline
SOTS &
$\begin{aligned}\widetilde{P}_{{out}}^\infty = & \sum_{\substack{n_1, n_2, \dots, n_{m_{{kd}}+1} \geq 0 \\ n_1 + n_2 + \dots + n_{m_{{kd}}+1} = N}} 
    \delta(N, \theta_1, \theta_2, \tilde{\lambda}_{kd}) \left( \frac{\tau \eta_2}{\eta_1} \right)^{\theta_2} \left( \frac{1}{\tilde\lambda_{{ke}} + \frac{\tilde\lambda_{{kd}} \theta_1 \tau \eta_2}{\eta_1}} \right)^{m_{{ke}} + \theta_2} 
    \Gamma(m_{{ke}} + \theta_2).
\end{aligned}$ \\ \hline
METS & $\begin{aligned} 
\widetilde{P}_{{out}}^\infty &= 1 - \sum_{j=0}^{m_{kd}-1} \left(\frac{\lambda_{kd}^j}{\Gamma(j+1)} \right) \sum_{\substack{n_1,n_2,\ldots,n_{m_{ke}}\geq0 \\ n_1+n_2+\cdots+n_{m_{ke}}=l}} 
\delta\left(l, \theta_3, \theta_4, \tilde{\lambda}_{ke}\right) \left(\frac{\tau \eta_2}{\eta_1}\right)^j \quad \left( \frac{1}{\lambda_{ke} \theta_3 + \frac{\lambda_{kd} \tau \eta_2}{\eta_1}} \right)^{\theta_4 + j} \nonumber \\
& \times \Bigg( \lambda_{ke} \theta_3 \left( \frac{1}{\lambda_{ke} \theta_3 + \frac{\lambda_{kd} \tau \eta_2}{\eta_1}} \right) \Gamma\left(\theta_4 + j + 1\right) \quad - \theta_4 \Gamma\left(\theta_4 + j\right) \Bigg).
\end{aligned}$ \\ \hline
OTS & $\begin{aligned}
\widetilde{P}_{{out}}^\infty &=  \left( 1 - \left( \frac{\tilde{\lambda}_{ke}^{m_{ke}}}{\Gamma(m_{ke})} \right) 
\sum_{j=0}^{m_{kd}-1} \frac{\tilde{\lambda}_{kd}^j}{\Gamma(j+1)} \left( \frac{\tau \eta_2}{\eta_1} \right)^j \right. \left. \left( \frac{1}{\tilde{\lambda}_{ke} + \frac{\tilde{\lambda}_{kd} \tau \eta_2}{\eta_1}} \right)^{m_{ke}+j} 
\Gamma(m_{ke}+j) \right)^N.
\end{aligned}$ \\ \hline
RTS & $\begin{aligned}
\widetilde{P}_{{out}}^\infty &=  \frac{1}{N} \sum_{n=1}^{N} \left( 1 - \left( \frac{\tilde{\lambda}_{ke}^{m_{ke}}}{\Gamma(m_{ke})} \right) 
\sum_{j=0}^{m_{kd}-1} \frac{\tilde{\lambda}_{kd}^j}{\Gamma(j+1)} \left( \frac{\tau \eta_2}{\eta_1} \right)^j \right. \left. \left( \frac{1}{\tilde{\lambda}_{ke} + \frac{\tilde{\lambda}_{kd} \tau \eta_2}{\eta_1}} \right)^{m_{ke}+j} 
\Gamma(m_{ke}+j) \right).
\end{aligned}$ \\ \hline
\end{tabular}
}
\end{table*}
% \vspace{-0.5cm}
%%%% Asymptotic IP
%\subsubsection{Asymptotic \ac{ip}}
\subsubsection{Asymptotic \texorpdfstring{\ac{ip}}{IP}}
The asymptotic \ac{ip} of any protocol can write as
\begin{align}
\label{asyip}
\widetilde{P}_{ip}^{\infty}&=\widetilde{P}_{1}^{\infty}+\widetilde{P}_{3}^{\infty},
% \nonumber\\&= \underbrace{\mathbb{P}\left(\beta_k^\ast=0\right)}_{=\widetilde{P}_1} + \underbrace{\mathbb{P}\left(\beta_k^\ast>0, \ \gamma_d^{k}<\gamma_e^k\right)}_{=\widetilde{P}_3}
\end{align}
where 
\begin{align}
\label{asyipP3}
&\widetilde{P}_{3}^{\infty}=\lim_{\Gamma_t\rightarrow \infty}\widetilde{P}_{3}
% \nonumber\\
% &
=\lim_{\Gamma_t\,\mbox{or}\,\mathcal{P}\rightarrow \infty}
\mathbb{P}\left(W_1>0,W_2<\frac{\eta_2 w_3}{\mathrm{\eta_1}}\right)\nonumber 
\\&=\int_{0}^{\infty}\int_{0}^{\infty}F_{W_2}\left(\frac{\eta_2 w_3}{\eta_1}\right) f_{W_1}\left(w_1\right)
% \nonumber\\&\times 
f_{W_3}\left(w_3\right)dw_1 dw_3.
\end{align}
The asymptotic \ac{ip} for each protocol is derived by solving (\ref{asyipP3}) using similar mathematical techniques as those applied for the exact \ac{ip} expressions. The derived asymptotic \ac{ip} formulations for all protocols are presented in Table~\ref{asyiptable} on page 14.
%%%% Asymptotic IP Table-IV
\begin{table*}[t!]
\centering
  \caption{Asymptotic IP Expressions}
   \label{asyiptable}
% \resizebox{\columnwidth}{!}{%
\scalebox{0.85}{
\begin{tabular}{|l|l|}
\hline
Protocol &  \multicolumn{1}{c|}{Asymptotic IP}\\ \hline
SOTS &
$\begin{aligned}\widetilde{P}_{{ip}}^\infty &=  \sum_{\substack{n_1, n_2, \dots, n_{m_{{kd}}+1} \geq 0 \\ n_1 + n_2 + \dots + n_{m_{{kd}}+1} = N}} 
    \delta(N, \theta_1, \theta_2, \tilde{\lambda}_{kd}) \left( \frac{\eta_2}{\eta_1} \right)^{\theta_2} \left( \frac{1}{\tilde\lambda_{{ke}} + \frac{\tilde\lambda_{{kd}} \theta_1 \eta_2}{\eta_1}} \right)^{m_{{ke}} + \theta_2} 
    \Gamma(m_{{ke}} + \theta_2).
\end{aligned}$ \\ \hline
METS & $\begin{aligned} 
\widetilde{P}_{{ip}}^\infty &= 1 - \sum_{j=0}^{m_{kd}-1} \left(\frac{\lambda_{kd}^j}{\Gamma(j+1)} \right) \sum_{\substack{n_1,n_2,\ldots,n_{m_{ke}}\geq0 \\ n_1+n_2+\cdots+n_{m_{ke}}=l}} 
\delta\left(l, \theta_3, \theta_4, \tilde{\lambda}_{ke}\right) \left(\frac{\eta_2}{\eta_1}\right)^j \quad \left( \frac{1}{\lambda_{ke} \theta_3 + \frac{\lambda_{kd}  \eta_2}{\eta_1}} \right)^{\theta_4 + j} \nonumber \\
& \times \Bigg( \lambda_{ke} \theta_3 \left( \frac{1}{\lambda_{ke} \theta_3 + \frac{\lambda_{kd} \eta_2}{\eta_1}} \right) \Gamma\left(\theta_4 + j + 1\right) \quad - \theta_4 \Gamma\left(\theta_4 + j\right) \Bigg).
\end{aligned}$ \\ \hline
OTS & $\begin{aligned}
\widetilde{P}_{{ip}}^\infty &=  \left( 1 - \left( \frac{\tilde{\lambda}_{ke}^{m_{ke}}}{\Gamma(m_{ke})} \right) 
\sum_{j=0}^{m_{kd}-1} \frac{\tilde{\lambda}_{kd}^j}{\Gamma(j+1)} \left( \frac{\eta_2}{\eta_1} \right)^j \right. \left. \left( \frac{1}{\tilde{\lambda}_{ke} + \frac{\tilde{\lambda}_{kd} \eta_2}{\eta_1}} \right)^{m_{ke}+j} 
\Gamma(m_{ke}+j) \right)^N.
\end{aligned}$ \\ \hline
RTS & $\begin{aligned}
\widetilde{P}_{{ip}}^\infty &=  \frac{1}{N} \sum_{n=1}^{N} \left( 1 - \left( \frac{\tilde{\lambda}_{ke}^{m_{ke}}}{\Gamma(m_{ke})} \right) 
\sum_{j=0}^{m_{kd}-1} \frac{\tilde{\lambda}_{kd}^j}{\Gamma(j+1)} \left( \frac{\eta_2}{\eta_1} \right)^j \right. \left. \left( \frac{1}{\tilde{\lambda}_{ke} + \frac{\tilde{\lambda}_{kd} \eta_2}{\eta_1}} \right)^{m_{ke}+j} 
\Gamma(m_{ke}+j) \right).
\end{aligned}$ \\ \hline
\end{tabular}
}
\end{table*}

%%%% NUMERICAL AND SIMULATION RESULTS
\section{NUMERICAL AND SIMULATION RESULTS}
\label{numerical_results}
 Here, we analyze the \ac{sop} and \ac{ip} performance of energy-harvesting \ac{backcom} systems with tag selection through numerical evaluations. The accuracy of the derived expressions is further confirmed using Monte Carlo simulations. The parameters related to considered \ac{backcom} network are chosen as:  $P_{max}^{T_k}=200\mu W$, $\xi_0=5\mu$W, $\xi_1=5000$, $\xi_2=0.0002$ and $P_{c}^{T_k}=200\mu$W. The channel parameters are set to $\tilde{\lambda}_{sk}=2$ dB, $\tilde{\lambda}_{kd}=3$ dB and $\tilde{\lambda}_{ke}=5$ dB. We assume 
 $d'_{sk}=d'_s$, and $d'_{kx}=d'_x$ for $k \in \{1,2,...,N\}$, with a path loss exponent $u_{sk}=u_{kx}=2$, $\Gamma_p=5$ dB, and a tag backscattering coefficient of 2.2. The data rate $\mathcal{R}$ is measured in bits/s/Hz.
 
\begin{figure*}[t!]
          \centering
          \subfigure[]
          {
          \includegraphics[width=5.1cm,height=4.5cm]{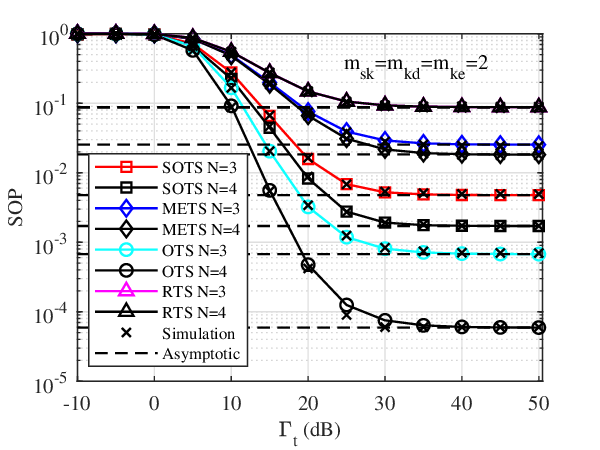}
          \label{fig2}
          }
          \subfigure[]
          {
          \includegraphics[width=5.1cm,height=4.5cm]{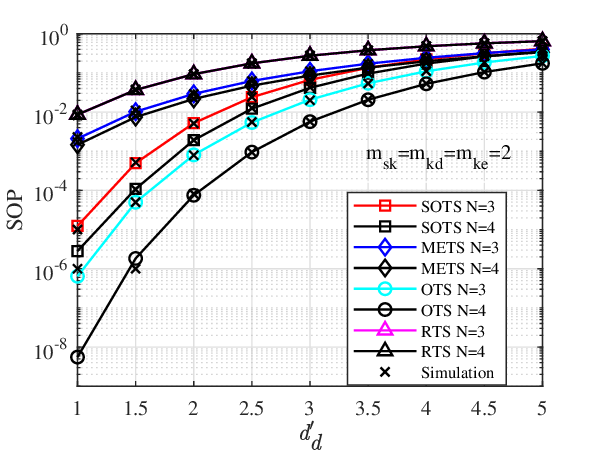}
         \label{fig3}
         }
         \subfigure[]
         {
          \includegraphics[width=5.1cm,height=4.5cm]{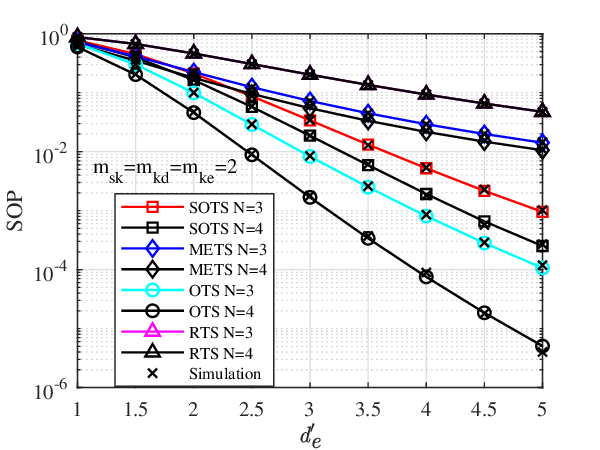}
          \label{fig4}
          }
        \caption{(a): SOP versus \(\Gamma_t\) for $N \in \{3,4\}$, (b): SOP versus $d'_d$ for $N \in \{3,4\}$, (c): SOP versus $d'_e$ for $N \in \{3,4\}$. }
        % \vspace*{-6mm}
\end{figure*}

 %%% Fig.-2
 Fig.~\ref{fig2} illustrates the \ac{sop} as a function of $\Gamma_t$ for different tag selection schemes with $N \in \{3,4\}$, using the parameters $d'_s = 1m$, $d'_d = 2m$, $d'_e = 4m$, $\mathcal{R} = 0.5$, and $m_{sk}=m_{kd}=m_{ke}=2$. The results show that the asymptotic \ac{sop} closely matches the analytical results at high \ac{snr}, validating the accuracy of the derived expressions. As $\Gamma_t$ and the number of tags $N$ increase, the \ac{sop} decreases across all schemes, demonstrating that higher transmission power and more selection options improve system secrecy. In terms of performance ranking, the schemes are ordered as \ac{ots}, \ac{sots}, \ac{mets}, and \ac{rts}, from best to worst. \ac{ots} achieves the lowest \ac{sop} because it optimally balances the trade-off between maximizing the desired signal strength and minimizing eavesdropper interference. \ac{sots} follows closely due to its efficient selection based on maximizing power gain while maintaining lower complexity. \ac{mets} provides moderate performance, as it prioritizes minimizing the strength of the eavesdropper's link but lacks optimality in other aspects. Lastly, \ac{rts} shows the highest SOP because it selects tags without considering the channel state information, making it the least effective in ensuring secure communications.

%%%%%%%%%%%%%%%% New figures 2-3-4
 
%%% Fig.-3
Fig.~\ref{fig3} examines the impact of the $\mathcal{T}_k-D$ distance on the secrecy performance of different tag selection schemes. The analysis uses the parameters $\Gamma_t = 30$ dB, $d'_s = 1m$, $d'_e = 4m$, $\mathcal{R} = 0.5$, $m_{sk}=m_{kd}=m_{ke}=2$, plotting \ac{sop} as a function of distance for varying $k$ values. The results demonstrate that as the distance between $\mathcal{T}_k$ and $D$ increases, the \ac{sop} also rises due to the greater path loss, which weakens the legitimate communication link relative to the eavesdropper's link. The performance gap between \ac{sots} and \ac{mets} decreases with larger $N$, as more tags improve the chances of selecting one with a favorable $\mathcal{T}_k-D$ link. \ac{ots} remains superior due to its optimal selection strategy, ensuring maximum secrecy capacity regardless of distance. \ac{rts} performs the worst, as its random selection ignores channel conditions, making it vulnerable to path loss and eavesdropping. This highlights the critical role of $\mathcal{T}_k-D$ distance in influencing the effectiveness of tag selection schemes for secure \ac{backcom} systems.

%%% Fig.-4
Fig.~\ref{fig4} illustrates the influence of $\mathcal{T}_k-E$ distance on secrecy performance, with \ac{sop} plotted against $d'_e$ for various $N$ and fixed $m_{sk}=m_{kd}=m_{ke}=2$. The results show that \ac{sop} improves as the eavesdropper moves farther from the tag, primarily due to increased path loss reducing the eavesdropper's received signal strength. This highlights the significance of spatial placement in enhancing the security of \ac{backcom} systems.

\begin{figure*}[t!]
          \centering
          \subfigure[]
          {
          \includegraphics[width=5.1cm,height=4.5cm]{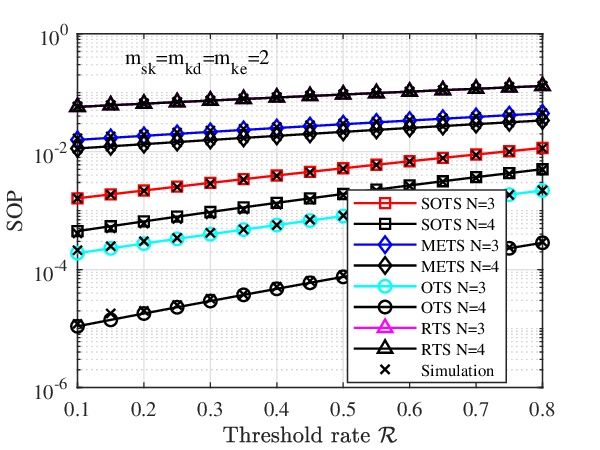}
          \label{fig5}
          }
          \subfigure[]
         {
          \includegraphics[width=5.1cm,height=4.5cm]{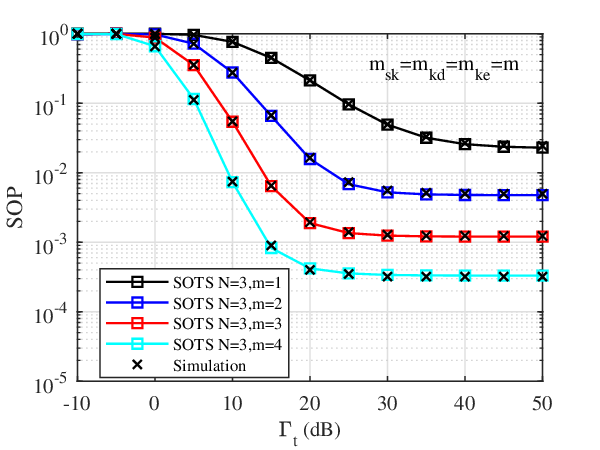}
          \label{fig6}
          }
          \subfigure[]
          {
          \includegraphics[width=5.1cm,height=4.5cm]{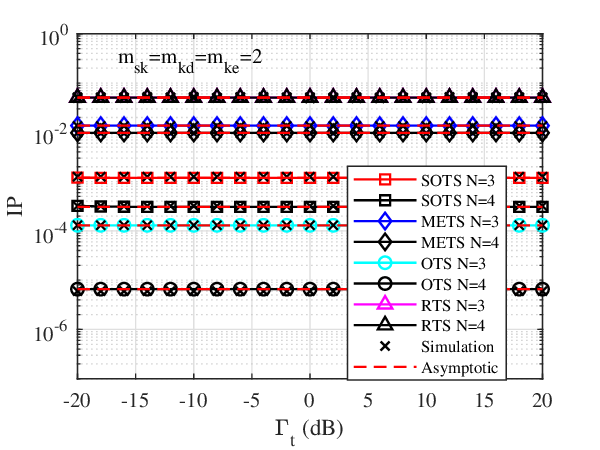}
         \label{fig7}
         }
         
        \caption{(a): SOP versus $\mathcal{R}$ for $N \in \{3,4\}$, (b): SOP versus \(\Gamma_t\) for $m \in \{1, 2, 3, 4\}$, (c): IP versus \(\Gamma_t\) for $N \in \{3,4\}$,. }
        % \vspace*{-6mm}
\end{figure*}

 %%% Fig.-5
Fig.~\ref{fig5} shows \ac{sop} as a function of the threshold rate $\mathcal{R}$ for various selection schemes with $N \in \{3,4\}$ and $m_{sk}=m_{kd}=m_{ke}=2$. SOP increases with higher threshold rates due to stricter secrecy requirements. The performance improves with more tags, as additional options enhance the likelihood of selecting a tag with favorable channel conditions. Among all schemes, \ac{ots} consistently achieves the best performance, with a more pronounced SOP improvement observed for 
$N=4$, showcasing the effectiveness of optimal tag selection in securing communication under tighter secrecy constraints.
%%%%%%%%%%%%%%%% New figures-5-6-7

% %%% Fig.-6
Fig.~\ref{fig6} illustrates the \ac{sop} as a function of $\Gamma_t$ for the \ac{sots} scheme with $N = 3$ and varying fading parameters $m$ ($m_{sk} = m_{kd} = m_{ke} = m$). The results reveal that as $m$ increases, the \ac{sop} improves due to reduced fading severity. This trend occurs because higher values of $m$ correspond to less severe channel fluctuations, resulting in more stable and reliable communication links. Consequently, increasing $m$ enhances the system's secrecy performance by reducing the probability of eavesdropping success. 

%%% Fig.-7
Fig.~\ref{fig7} illustrates \ac{ip} as a function of 
$\Gamma_t$ for various selection schemes with $N \in \{3,4\}$ and $m_{sk}=m_{kd}=m_{ke}=2$. Consistent with the trends observed for \ac{sop}, \ac{ip} shows improvement as $\Gamma_t$ and $N$ increase, primarily due to the expanded set of available tags, which enhances the chances of selecting one with favorable communication conditions.

%%%% Conclusion	
\section{Conclusion}
\label{conclusion}
This paper investigates the secrecy performance of \ac{eh} \ac{backcom} systems using various tag selection schemes-including \ac{sots}, \ac{mets}, \ac{ots}, and \ac{rts}-in Nakagami-$m$ fading channels. We derived closed-form expressions for \ac{sop} and \ac{ip} to compare all schemes. We also provided asymptotic expressions for \ac{sots}, \ac{mets}, \ac{ots}, and \ac{rts}, offering deeper insights into performance trade-offs. The performance comparison of these selection protocols in terms of \ac{sop} is as follows: \ac{ots} demonstrates the best performance, followed by \ac{sots}, \ac{mets}, and finally \ac{rts}. Nakagami-$m$ fading proves valuable in analyzing how different channel conditions impact secure communications. Our derived theoretical results verified by simulation results.

% %Appendix
% \appendices
\begin{appendix}
%%% Appendix-A for SOTS  
%\subsection{Derivation of $\widetilde{P}_2$ for SOTS}
\subsection{Derivation of \texorpdfstring{$\widetilde{P}_2$}{P2 (tilde)} for SOTS}
\label{sotsappendix}
Using binomial expansion and after simplifying, the $F_{W_2
}(\cdot)$ can be written as 
% (\ref{cdfW2b}) 
% on the next page.
% \begin{figure*}[t!]
% \normalsize
% \setcounter{equation}{37}
\begin{align}
\label{cdfW2b}
&F_{W_2}\left(\frac{\tau-1}{\beta_n^\ast\eta_1 \Gamma_t w _1}+\frac{\tau\eta_2 w_3}{\eta_1}\right)
=\sum_{\substack{n_1,n_2,\ldots,n_{m_{kd}+1}\geq 0 \\ n_1+n_2+\ldots+n_{m_{kd}+1}=N}}\delta\left(N, \theta_1,\theta_2,\tilde\lambda_{kd}\right) 
%%%
\sum_{q=0}^{\theta_2} \binom{\theta_2}{q} \left(\frac{\tau\eta_2}{\eta_1}\right)^q 
\nonumber 
\\ &  \times
\left(\frac{\tau - 1}{\eta_1 \Gamma_t \left(w_1 - \frac{\phi}{\mathcal{P}{d'^{\,-u_{sk}}_{sk}}}\right)}\right)^{\theta_2 - q} 
% \nonumber 
% \\ &  \times
\exp\left(-\frac{(\tau - 1)\tilde{\lambda}_{kd} \theta_1 }{\eta_1 \Gamma_t \left(w_1 - \frac{\phi}{\mathcal{P} d'^{-u_{sk}}_{sk}}\right)} \right)
% \nonumber 
% \\ &  \times
w_3^q \exp\left(-\frac{\tilde{\lambda}_{kd} \theta_1\tau \eta_2}{\eta_1}w_3\right).
\end{align}
% \hrulefill
% \end{figure*}
By substituting $F_{W_2}(\cdot)$ which is given in (\ref{cdfW2b}) into (\ref{P2sostb}), we can obtain
\begin{align}
\label{P2sotsc}
   \widetilde{P}_2&= \sum_{\substack{n_1,n_2,\ldots,n_{m_{kd}+1}\geq 0 \\ n_1+n_2+\ldots+n_{m_{kd}+1}=N}}\delta\left(N, \theta_1,\theta_2,\tilde\lambda_{kd}\right) 
%%%
\sum_{q=0}^{\theta_2} \binom{\theta_2}{q} 
% \nonumber\\&\times
\left(\frac{\tau\eta_2}{\eta_1}\right)^q \mathcal{I}_1 \mathcal{I}_2,
\end{align}
% On Solving $P_2$, we used the Multinomial Theorem (Proof: Refer to Appendix A) as
where 
\begin{align}
\label{I1a}
\mathcal{I}_1 &=\int_{\frac{\phi}{\mathcal{P}{d'^{\,-u_{sk}}_{sk}}}}^{\infty}\left(\frac{\tau - 1}{\eta_1 \Gamma_t \left(w_1 - \frac{\phi}{\mathcal{P}{d'^{\,-u_{sk}}_{sk}}}\right)}\right)^{\theta_2 - q} 
% \nonumber 
% \\ &  \times
\exp\left(-\frac{(\tau - 1)\tilde{\lambda}_{kd} \theta_1 }{\eta_1 \Gamma_t \left(w_1 - \frac{\phi}{\mathcal{P} d'^{-u_{sk}}_{sk}}\right)} \right)
 f_{W_1}(w_1)\, dw_1,
\end{align}
and 
\begin{align}
\label{I2a}
\mathcal{I}_2 &= \int_{0}^{\infty} w_3^q \exp\left(-\frac{\tilde{\lambda}_{kd} \theta_1\tau \eta_2}{\eta_1}w_3\right) f_{W_3}(w_3)\, dw_3. 
\end{align}
Substituting $f_{W_1}(\cdot)$ into (\ref{I1a}) and applying change of variable $v_1=w_1-\frac{\phi}{\mathcal{P}{d'^{\,-u_{sk}}_{sk}}}$, we can obtain
\begin{align}
\label{I1b}
\mathcal{I}_1 &=\exp\left(-\frac{\tilde{\lambda}_{sk}\phi}{\mathcal{P}{d'^{\,-u_{sk}}_{sk}}}\right)\left(\frac{\tau - 1}{\eta_1 \Gamma_t }\right)^{\theta_2 - q} \int_{0}^{\infty}
v_1^{q-\theta_2}
% \nonumber 
% \\ &  \times
\left(v_1+\frac{\phi}{\mathcal{P}{d'^{\,-u_{sk}}_{sk}}}\right)^{m_{sk}-1}
\nonumber 
\\ &  \times
\exp\left(-\left(\tilde{\lambda}_{sk}v_1+\frac{(\tau - 1)\tilde{\lambda}_{kd} \theta_1 }{\eta_1 \Gamma_t v_1} \right)\right)\, dv_1.
\end{align}
Furthermore, using binomial expansion for $\left(v_1+\frac{\phi}{\mathcal{P}{d'^{\,-u_{sk}}_{sk}}}\right)^{m_{sk}-1}$, we can get
\begin{align}
\label{I1c}
\mathcal{I}_1 &=\exp\left(-\frac{\tilde{\lambda}_{sk}\phi}{\mathcal{P}{d'^{\,-u_{sk}}_{sk}}}\right)\left(\frac{\tau - 1}{\eta_1 \Gamma_t }\right)^{\theta_2 - q}
\sum_{p=0}^{m_{sk} - 1} \binom{m_{sk} - 1}{p} 
% \nonumber 
% \\ &  \times 
\left( \frac{\phi}{\mathcal{P} d'^{-u_{sk}}_{sk}} \right)^{m_{sk} - 1 - p}
\nonumber 
\\ &  \times
\int_{0}^{\infty}
v_1^{p+q-\theta_2}
% \nonumber 
% \\ &  \times
\exp\left(-\left(\tilde{\lambda}_{sk}v_1+\frac{(\tau - 1)\tilde{\lambda}_{kd} \theta_1 }{\eta_1 \Gamma_t v_1} \right)\right)\, dv_1.
\end{align}
Using \cite[(2.3.16.1)]{prudnikov1986integralsvol1}, we can get a closed-form solution of $\mathcal{I}_1$ as
\begin{align}
\label{I1d}
\mathcal{I}_1 &=\exp\left(-\frac{\tilde{\lambda}_{sk}\phi}{\mathcal{P}{d'^{\,-u_{sk}}_{sk}}}\right)\left(\frac{\tau - 1}{\eta_1 \Gamma_t }\right)^{\theta_2 - q}
\sum_{p=0}^{m_{sk} - 1} \binom{m_{sk} - 1}{p} 
% \nonumber 
% \\ &  \times 
\left( \frac{\phi}{\mathcal{P} d'^{-u_{sk}}_{sk}} \right)^{m_{sk} - 1 - p}
\nonumber\\& \times
\left(2 \left( \frac{\tilde{\lambda}_{kd} \theta_1 (\tau - 1)}{\eta_1 \Gamma_t \tilde{\lambda}_{sk}} 
\right)^{\frac{p + q - \theta_2 + 1}{2}}
% \right.
% \nonumber\\& \left.\times
K_{p + q - \theta_2 + 1} \left( \sqrt{\frac{4 \theta_1 \tilde{\lambda}_{sk} \tilde{\lambda}_{kd} 
(\tau - 1)}{\eta_1 \Gamma_t}} \right) \right).
\end{align}
Substituting $f_{W_3}(\cdot)$ into (\ref{I2a}) and solving the integral using \cite[(8.310.1)]{gradshteyn2007table}, we can obtain a closed-form solution of $\mathcal{I}_2$ as
\begin{align}
\label{I2b}
\mathcal{I}_2 &= \left(\tilde{\lambda}_{ke} + \frac{\tilde{\lambda}_{kd} \theta_1 \tau \eta_2}{\eta_1} 
\right)^{-m_{ke} - q} \Gamma(m_{ke} + q). 
\end{align}
Using (\ref{I1d}) and (\ref{I2b}) into (\ref{P2sotsc}), we can get a new closed-form expression of $\widetilde{P}_2$ which is given in (\ref{P2sotsFinal}) on page 8. 

%%% Appendix for METS  
% \subsection{Derivation of $\widetilde{P}_2$ for METS}
\subsection{Derivation of \texorpdfstring{$\widetilde{P}_2$}{P2} for METS}
\label{metsappendix}
By substituting $F_{W_2}(\cdot)$ which is given in (\ref{cdfgkx}) into (\ref{P2metsb}), we can obtain
\begin{align}
\label{P2metsc}
   \widetilde{P}_2&=\mathcal{I}_3 -\sum_{j=0}^{m_{ke}-1}\frac{\tilde\lambda_{kd}^j}{\Gamma(j+1)}\mathcal{I}_4,
\end{align}
% On Solving $P_2$, we used the Multinomial Theorem (Proof: Refer to Appendix A) as
where 
\begin{align}
\label{I3a}
\mathcal{I}_3 &= \int_{0}^{\infty} \int_{\frac{\phi}{\mathcal{P}{d'^{\,-u_{sk}}_{sk}}}}^{\infty}
 f_{W_1}(w_1)f_{W_3}(w_3)\, dw_1dw_3,
\end{align}
and 
\begin{align}
\label{I4a}
\mathcal{I}_4 &= \int_{0}^{\infty} \int_{\frac{\phi}{\mathcal{P}{d'^{\,-u_{sk}}_{sk}}}}^{\infty}
\left(\frac{\tau - 1}{\eta_1 \Gamma_t \beta_k^{\ast} w_1 }+\frac{ w_3}{\eta_1}\right)^{j} 
% \nonumber 
% \\ &  \times
\exp\left(-\tilde{\lambda}_{kd}\left(\frac{\tau - 1}{\eta_1 \Gamma_t \beta_k^{\ast} w_1 }+\frac{ w_3}{\eta_1}\right) \right)
\nonumber 
\\ &  \times 
f_{W_1}(w_1)f_{W_3}(w_3)\, dw_1dw_3. 
\end{align}
Using properties of \ac{pdf} and \ac{cdf}, we can obtain a closed-form solution of $\mathcal{I}_3$ as
\begin{align}
\label{I3b}
\mathcal{I}_3 &= \int_{0}^{\infty} f_{W_3}(w_3)\, dw_3\int_{\frac{\phi}{\mathcal{P}{d'^{\,-u_{sk}}_{sk}}}}^{\infty}
 f_{W_1}(w_1)\, dw_1
 % \nonumber\\&
 % =1-F_{W_1}\left(\frac{\phi}{\mathcal{P}{d'^{\,-u_{sk}}_{sk}}}\right)
 =1-\frac{\gamma\left(m_{sk},\frac{\tilde\lambda_{sk}\phi}{\mathcal{P}{d'^{\,-u_{sk}}_{sk}}}\right)}{\Gamma\left(m_{sk}\right)}.
\end{align}
Using binomial expansion for $\left(\frac{\tau-1}{\eta_1 \Gamma_t \beta_k^{\ast} w_1 }+\frac{ w_3}{\eta_1}\right)^{j} $ and simplification, we can write 
\begin{align}
\label{I4b}
\mathcal{I}_4&=\sum_{q=0}^{j}\binom{j}{q} \left(\frac{\tau \eta_2}{\eta_1}\right)^q \mathcal{I}_{41}\mathcal{I}_{42},
\end{align}
where 
\begin{align}
\label{I41a}
\mathcal{I}_{41}&=\int_{\frac{\phi}{\mathcal{P}{d'^{\,-u_{sk}}_{sk}}}}^{\infty}\left(\frac{\tau - 1}{\eta_1 \Gamma_t \left(w_1 - \frac{\phi}{\mathcal{P}{d'^{\,-u_{sk}}_{sk}}}\right)}\right)^{j - q} 
% \nonumber 
% \\ &  \times
\exp\left(-\frac{(\tau - 1)\tilde{\lambda}_{kd} }{\eta_1 \Gamma_t \left(w_1 - \frac{\phi}{\mathcal{P} d'^{-u_{sk}}_{sk}}\right)} \right)
 f_{W_1}(w_1)\, dw_1,
\end{align}
and 
\begin{align}
\label{I42a}
\mathcal{I}_{42}&=\int_{0}^{\infty} w_3^q \exp\left(-\frac{\tilde{\lambda}_{kd} \theta_1\tau \eta_2}{\eta_1}w_3\right) f_{W_3}(w_3)\, dw_3.
\end{align}
Substituting $f_{W_1}(\cdot)$ into (\ref{I41a}) and applying change of variable $v_1=w_1-\frac{\phi}{\mathcal{P}{d'^{\,-u_{sk}}_{sk}}}$, we can obtain as
% (\ref{I41b}) on the next page.
%%%
% \begin{figure*}[t!]
% \normalsize
% \setcounter{equation}{52}
\begin{align}
\label{I41b}
\mathcal{I}_{41}&=\left(\frac{\tau - 1}{\eta_1 \Gamma_t }\right)^{j-q}\frac{(\tilde{\lambda}_{sk}^{m_{sk}}}{\Gamma(m_{sk})}\exp\left(-\frac{\tilde{\lambda}_{sk}\phi}{\mathcal{P}{d'^{\,-u_{sk}}_{sk}}}\right) 
% \nonumber 
% \\ &  \times
\int_{0}^{\infty}
v_1^{q-j}
\left(v_1+\frac{\phi}{\mathcal{P}{d'^{\,-u_{sk}}_{sk}}}\right)^{m_{sk}-1}
\nonumber 
\\ &  \times
\exp\left(-\left(\tilde{\lambda}_{sk}v_1+\frac{(\tau - 1)\tilde{\lambda}_{kd}}{\eta_1 \Gamma_t v_1} \right)\right)\, dv_1.
\end{align}
% \hrulefill
% \end{figure*}
Using binomial expansion for $\left(v_1+\frac{\phi}{\mathcal{P}{d'^{\,-u_{sk}}_{sk}}}\right)^{m_{sk}-1}$ in (\ref{I41b}), we can get as
% (\ref{I41c}) on the next page.
%%%
% \begin{figure*}[t!]
% \normalsize
% \setcounter{equation}{53}
\begin{align}
\label{I41c}
\mathcal{I}_{41} &=\left(\frac{\tau - 1}{\eta_1 \Gamma_t }\right)^{j-q}\frac{(\tilde{\lambda}_{sk}^{m_{sk}}}{\Gamma(m_{sk})}\exp\left(-\frac{\tilde{\lambda}_{sk}\phi}{\mathcal{P}{d'^{\,-u_{sk}}_{sk}}}\right) 
% \nonumber 
% \\ &  \times
\sum_{p=0}^{m_{sk} - 1} \binom{m_{sk} - 1}{p}  \left( \frac{\phi}{\mathcal{P} d'^{-u_{sk}}_{sk}} \right)^{m_{sk} - 1 - p}
\int_{0}^{\infty}
v_1^{p+q-k}
\nonumber 
\\ &  \times
\exp\left(-\left(\tilde{\lambda}_{sk}v_1+\frac{(\tau - 1)\tilde{\lambda}_{kd}}{\eta_1 \Gamma_t v_1} \right)\right)\, dv_1.
\end{align}
% \hrulefill
% \end{figure*}
% \vspace{-1cm}
Using \cite[(2.3.16.1)]{prudnikov1986integralsvol1}, we can get a closed-form solution of $\mathcal{I}_{41}$ as  
% (\ref{I41d}) on the next page.
%%%%%%%%%%%%===
% \begin{figure*}[t!]
% \normalsize
% \setcounter{equation}{54}
\begin{align}
\label{I41d}
\mathcal{I}_{41} &=\left(\frac{\tau - 1}{\eta_1 \Gamma_t }\right)^{j-q}\frac{(\tilde{\lambda}_{sk}^{m_{sk}}}{\Gamma(m_{sk})}\exp\left(-\frac{\tilde{\lambda}_{sk}\phi}{\mathcal{P}{d'^{\,-u_{sk}}_{sk}}}\right) 
% \nonumber 
% \\ &  \times
\sum_{p=0}^{m_{sk} - 1} \binom{m_{sk} - 1}{p}  \left( \frac{\phi}{\mathcal{P} d'^{-u_{sk}}_{sk}} \right)^{m_{sk} - 1 - p}
\nonumber\\& \times
\left(2 \left( \frac{\tilde{\lambda}_{kd} (\tau - 1)}{\eta_1 \Gamma_t \tilde{\lambda}_{sk}} 
\right)^{\frac{p + q -j + 1}{2}}
% \right.
% \nonumber\\& \left.\times
K_{p + q -j + 1} \left( \sqrt{\frac{4\tilde{\lambda}_{sk} \tilde{\lambda}_{kd} 
(\tau - 1)}{\eta_1 \Gamma_t}} \right) \right).
\end{align}
% \hrulefill
% \end{figure*}
%%%
Substituting $f_{W_3}(\cdot)$ into (\ref{I42a}) and solving the integral using \cite[(8.310.1)]{gradshteyn2007table}, we can obtain a closed-form solution of $\mathcal{I}_{42}$ as 
% (\ref{I42b}) on the next page. 
% \begin{figure*}[htb!]
% \normalsize
% \setcounter{equation}{55}
\begin{align}
\label{I42b}
\mathcal{I}_{42} &=  \sum_{l=1}^{N}\binom{N}{l}(-1)^{l+1} 
% \nonumber\\
% & \times
\sum_{\substack{n_1,n_2,\ldots,n_{m_{ke}}\geq0 \\ n_1+n_2+\cdots+n_{m_{ke}}=l}} 
\delta\left(l, \theta_3, \theta_4, \tilde{\lambda}_{ke}\right)  \left(\tilde\lambda_{ke} \theta_3 + \frac{\tilde\lambda_{kd} \tau \eta_2}{\eta_1} \right)^{-(\theta_4+q)} \Gamma(\theta_4+q+1)
\nonumber\\
& \times
\left( \frac{\tilde\lambda_{{ke}} \theta_3}{\tilde\lambda_{{ke}} \theta_3 + \frac{\tilde\lambda_{{kd}} \tau \eta_2}{\eta_1}} 
- \frac{\theta_4}{\theta_4 + q} \right). 
\end{align}
% \hrulefill
% \end{figure*}
Using (\ref{I3b}) and (\ref{I4b}) into (\ref{P2metsc}), we can get a new closed-form expression of $\widetilde{P}_2$ which is given in (\ref{P2metsFinal}) on page 10. 
\end{appendix}

%%%
 \bibliographystyle{IEEEtran}
	\footnotesize
	\bibliography{References}

\end{document}